\documentclass[twocolumn, aps, pre]{revtex4}

\usepackage{amsmath}
\usepackage{amsfonts}
\usepackage{amssymb}
\usepackage{graphicx}
\usepackage{pgfplots}
\usepackage{float}
\usepackage[caption=false]{subfig}
\usepackage{mathtools}
\usepackage{xcolor}
\usepackage{physics}
\usepackage{ragged2e}
\usepackage{float}
\PassOptionsToPackage{breaklinks}{hyperref}
\usepackage{hyperref}
\usepackage{enumerate}
\newcommand{\angular}[1]{
  \left\langle
    #1
  \right\rangle
}

\newcommand{\rhobar}{\bar{\rho}}
\newcommand{\intc}{\mathcal{Q}}
\newcommand{\hata}{\hat{a}}

\newcommand{\jfl}{\mathcal{J}^{(fl)}}
\newcommand{\jd}{\mathcal{J}^{(d)}}

\begin{document}

\title{Dynamic correlations in the conserved Manna sandpile}
\author{Anirban Mukherjee} 
\author{Punyabrata Pradhan}
 \affiliation{Department of Physics of Complex Systems, S. N. Bose
   National Centre for Basic Sciences, Block-JD, Sector-III, Salt Lake,
   Kolkata 700106, India.}

\begin{abstract}

We study dynamic correlations for current and mass, as well as the associated power spectra, in the one-dimensional conserved Manna sandpile. We show that, in the thermodynamic limit, the variance of cumulative bond current up to time $T$ grows subdiffusively as $T^{1/2-\mu}$ with the exponent $\mu \ge 0$ depending on the density regimes considered and, likewise, the power spectra of current and mass at low frequency $f$ varies as $f^{1/2+\mu}$ and $f^{-3/2+\mu}$, respectively; our theory predicts that, far from criticality, $\mu = 0$ and, near criticality, $\mu = (\beta+1)/2 \nu_{\perp} z > 0$ with $\beta$,  $\nu_{\perp}$ and $z$ being the order-parameter, correlation-length and dynamic exponents, respectively.
 The anomalous suppression of fluctuations near criticality signifies a ``dynamic hyperuniformity'', characterized by a set of fluctuation relations, in which current, mass and tagged-particle displacement fluctuations are shown to have a precise quantitative relationship with the density-dependent activity (or, its derivative). 
 In particular, the relation, ${\cal D}_s(\bar{\rho}) = a(\rhobar) / \rhobar$, between the self-diffusion coefficient ${\cal  D}_s(\bar{\rho})$, activity $a(\bar{\rho})$ and density $\rhobar$ explains a previous simulation observation [Eur. Phys. J. B \textbf{72}, 441 (2009)] that, near criticality, the self-diffusion coefficient in the Manna sandpile has the same scaling behavior as the activity.

\end{abstract}

\maketitle

\section{Introduction}
\label{sec:intro}

Long-ranged temporal correlations are ubiquitous in nature
\cite{Mandelbrot_83}. They usually manifest as the so-called ``$1/f$" or the
flicker noise, having a characteristic low-frequency power spectra
with a power-law form $1/f^{\psi}$, where  $0 < \psi < 2$, over a wide
range of frequency $f$. In fact, the $1/f$ noise has been seen in a variety of seemingly unrelated systems, such as solar flares
\cite{Frontera_1979, Aschwanden2014}, forest fires
\cite{Palmieri2020}, electrical activities in brain
\cite{Novikov_1997, hesse2014}, stock-market fluctuations
\cite{Tebaldi2021}, water flow in rivers \cite{Mandelbrot_83} and
resistance fluctuations in conductors \cite{Voss1976}, among others;
for details, see reviews \cite{jensen1998, bak_97}. However there is no general theory explaining the relative abundance of  $1/f$ noise in nature.

Bak, Tang, and Wisenfeld (BTW) proposed sandpiles as paradigmatic models of "self-organized criticality" (SOC) in order to provide a generic mechanism for long-ranged correlations in natural systems, and $1/f$ noise in particular. \cite{Bak_1987, Bak_1988}.
Sandpiles are spatially extended and threshold-activated systems, in which dynamical activities spread through cascades of toppling events (initialed when a local threshold is crossed), resulting in ``avalanche''-like dynamical activities and long-ranged correlations in the systems. They were envisaged as model-systems driven by slow addition of ``energy'', or grains, with local energy conservation in the bulk and dissipation at the boundary. Due to an intriguing interplay between drive and dissipation, the system evolves, apparently without fine-tuning of any parameters, towards a nonequilibrium steady state characterized through avalanches at all scales, i.e., a scale-invariant critical state with power-law distributions concerning various observables. Later, several variants, known as the conserved or {\it ``fixed-energy''} sandpiles, were proposed, where there is no dissipation, but the total number of grains (or, total energy) remains conserved, thus allowing the critical state to be reached by tuning the global density.

There have been numerous studies of sandpiles, and significant progress has been made in characterizing the static and dynamic properties of both critical and off-critical states of the systems. However, apart from some exact \cite{Dhar_1989, Dhar_1990, Priezzhev_1996, Majumdar_1992} and mathematically rigorous \cite{Fey_2010, Rolla_2021, Hazra_2017} results, the majority of the studies have been carried out using simulations \cite{Dickman_1998, Dickman_2001, Vespignani_2000} and phenomenological field-theoretical descriptions \cite{Vespignani_1998_II, Doussal_2015, Munoz_2004}. This is primarily due to the fact that the steady-state measure of a driven interacting-particle system such as sandpiles is in most cases a-priori unknown and, as a result, analytic calculations, beginning with a microscopic dynamical description, prove to be quite challenging \cite{Dickman_cluster_2002, Doussal_2015, Cunha_2009}.         
Perhaps not surprisingly, despite the fact that an explanation of $1/f$ noise was the main motivation for BTW's introduction of sandpile models, a good theoretical understanding of their time-dependent properties, particularly the exact hydrodynamics and the related transport coefficients  governing the large-scale relaxations, remains lacking ~\cite{Dhar_2006, Dickman_2000, redig_2006, Pradhan_2021, Ali_1995}.
Another fascinating aspect of sandpiles has only recently been discovered.  That is, the critical state of sandpiles can be {\it hyperuniform} \cite{Levine_PRL1_2015, Levine_PRL2_2017}, meaning that the static sub-system mass fluctuations scale with sub-system size in an anomalously slow manner \cite{Torquato_2003, Torquato_2016}. However, we currently have a limited knowledge of hyperuniform states of matter, and one can only speculate as to how such a state emerges dynamically in the first place \cite{Ilday_2021}. In fact, except for a few exact results concerning static properties of the hyperuniform state \cite{Dandekar-Dhar_2013, Dandekar_2020}, there has been little theoretical progress in this direction and a systematic approach for identifying the precise microscopic dynamical origin of the anomalous fluctuations remains elusive. In this scenario, a closer examination of the underlying dynamical mechanism that results in such a state is desirable.

Here we address the above issues in the context of conserved stochastic sandpiles and specifically focus on a continuous-time variant of the celebrated Manna sandpile \cite{Manna_1991, Dickman_2001}, which has drawn a lot of attention in the past \cite{Dickman_2000}.
The conserved Manna sandpile is a paradigm for systems, exhibiting a nonequilibrium {\it absorbing phase transition} from a dynamically active state to an absorbing one having no activities, upon tuning the global density. In fact, through several simulation studies in the recent past, it is known that the critical state of the conserved Manna sandpile is indeed hyperuniform \cite{Levine_PRL1_2015}. In another recent studies, it has been shown that the (near-)critical state is characterized by the singular transport coefficients, leading to anomalous relaxation and particle-transport in the system \cite{Chatterjee_PRE2018, Tapader_PRE2021}.

Historically, time-dependent properties of sandpiles have been studied in terms of power spectra of dynamical activity such as instantaneous toppling events in the systems \cite{Kertesz_1990, Laurson_2005}. In the original slowly-driven version of sandpiles, BTW had reported $1/f^{\psi}$ power-law behavior of the power spectrum for the activity, with the exponent $\psi < 2 $ \cite{Bak_1987}, although their claim was refuted as several simulation studies later found the exponent $\psi=2$ \cite{Jensen_1989,Kertesz_1990}. Subsequently, a more careful scaling analysis of  simulation data however revealed a nontrivial power spectra with the exponent $\psi < 2$ ~\cite{Laurson_2005}.
On the theoretical front, a dynamic renormalization-group analysis of phenomenological field-theoretical equations describing a ``running'' sandpile (driven with nonzero rate of grain addition) allowed an analytical calculation of the exponents $\psi$, involving activity as well as output current, where $1/f$-type noise was observed, with $\psi=-1$; quite interestingly, the temporal correlations in the long-time regime  were found to be {\it anti-correlated}, with $\psi=-1 < 0$.
This could well be  the earliest signature of {\it ``dynamic
  hyperuniformity''} in sandpiles. Recently, this particular aspect of
dynamic hyperuniformity, i.e., hyperuniformity in the temporal domain,
was also analyzed  in a variant of the slowly-driven
sandpiles, called the Oslo ricepile \cite{Garcia-Millan_2018}. Similar
low-frequency behavior of the activity power spectrum with $\psi=-1$
(i.e., anti-correlated) was observed in Ref.~\cite{Yadav_2012} for a
directed deterministic sandpile on a ladder with a finite driving
rate; in the slow-driving limit though,  Maslov {\it et
  al.}~\cite{Maslov_1999} previously showed the model to exhibit
$1/f^{\psi}$ power spectrum, with $\psi=1$, for the total mass
fluctuation. In a slightly different study~\cite{Maes_EPL2006}
of a driven sandpile, albeit on a periodic domain, the power-spectrum of activity had been found to be $1/f^\psi$ with $\psi=1$.
In a conserved deterministic lattice gas in two
dimensions~\cite{Jensen_PRE2012}, subsystem mass fluctuation was found
to exhibit the power spectrum $S_M(f) \sim f^{-\psi_M}$, where $\psi =
1.5$ away from criticality and $\psi \approx 1.8$ near
criticality.


In this paper, we theoretically investigate the time-dependent
correlations for current and mass in the (quasi-)steady state of the
one-dimensional conserved Manna sandpile.
We begin with a microscopic dynamical description of the model and then introduce a new, albeit approximate, closure scheme that allows us to analytically calculate the time-dependent correlation functions for current and mass, as well as the corresponding power spectra.
We establish a direct
quantitative relationship between various static and dynamic
fluctuation properties in terms of the density-dependent activity -
the system's ``order parameter'', and its derivative.
The main results of our paper are summarized as following.

\begin{enumerate}[(1)]
\item \textit{Time-integrated bond current fluctuation:}
  We show that, in the thermodynamic limit, with system size $L
  \rightarrow \infty$ and density $\rhobar$ fixed, the variance of the
  local (bond) current ${\mathcal{Q}(T)}$ up to time $T$ grows
  subdiffusively with time. That is, we have
  $\angular{\mathcal{Q}^2(T)} \sim T^{\alpha}$, where, away from
  criticality (in the time regime $T \ll L^2$), the exponent $\alpha =
  1/2$ and, near criticality (in the regime $T \ll L^z$), the current
  fluctuation is further suppressed with the exponent $\alpha=1/2
  -\mu$, where $\mu = (\beta+1)/2 \nu_{\perp} z > 0$ and $\beta$,
  $\nu_{\perp}$ and $z$ are the activity, correlation-length and
  dynamic exponents, respectively; thus the anomalous suppression of
  the current fluctuation near criticality serves as the {\it dynamic
    precursor} to the hyperuniform state formed at the critical point.

\item \textit{Power spectrum of current:}
  We find that the time-dependent (two-point) correlation function for
  the instantaneous current is long-ranged (power-law) and negative,
  resulting in 
  the low-frequency behavior of the corresponding power spectrum
  $S_{\cal J}(f) \sim f^{\psi_{\mathcal{J}}}$, which vanishes at low
  frequency, where $\psi_{\mathcal{J}} = 1/2$ away from criticality
  (strictly speaking, in the time regime $1/L^2 \ll f \ll 1$ for
  finite $L$) and $\psi_{\mathcal{J}} = 1/2 + \mu$ near criticality
  (in the time regime $1/L^z \ll f \ll 1$).

\item \textit{Power spectrum of mass:}
  We show that the power spectrum $S_{M}(f)$ for subsystem-mass
fluctuation on the other hand diverges $S_{M}(f) \sim f^{-\psi_M}$ at
low frequency, where $\psi_M = 3/2$ away from criticality ($1/L^2 \ll
f \ll 1$) and $\psi_M =3/2 - \mu$ near criticality ($1/L^z \ll f \ll
1$). These two exponents are not independent though and they are
connected by a scaling relation $\psi_M = 2 - \psi_{\mathcal{J}}$.

\item \textit{Time-integrated subsystem current fluctuation:}
  In the opposite limit of $T \rightarrow \infty$ (more
  specifically, $T \gg L^2$), with system size $L$ finite (but, still
  large) and density $\rhobar$ fixed, the scaled current fluctuation
  in the steady state is shown to be proportional to the activity
  $a(\rhobar)$ as $\lim_{T \rightarrow \infty}
  \angular{\mathcal{Q}^2(T)}/T = 2 a(\rhobar)/L$. On the other hand,
  the steady-state fluctuation of the subsystem current
  $\bar{Q}(l,T)$, i.e., the cumulative (summed over bonds) current in
  a subsystem of size $l$,  in the thermodynamic limit ($L \rightarrow
  \infty$) interestingly depends on the order of the limits
  taken. When the infinite-subsystem-size limit is taken first and
  then the infinite-time limit, the scaled subsystem current
  fluctuation $\sigma_Q^2(\rhobar) \equiv \lim_{l \rightarrow \infty}
  \lim_{T \rightarrow \infty} \langle \bar{Q}^2(l,T) \rangle/l T = 2
  a(\rhobar)$ converges to twice the activity.

\item \textit{Subsystem mass fluctuation and a fluctuation relation:}
  We derive a nonequilibrium fluctuation relation, $\sigma^2(\rhobar) =
\sigma_Q^2(\rhobar)/2D(\rhobar)$, which connects the (scaled)
subsystem mass fluctuation $\sigma^2(\rhobar) = \langle M_l^2 \rangle
- \langle M_l \rangle^2$, with $M_l$ being mass in a subsystem of size
$l$ and $D(\rhobar)$ the density-dependent bulk-diffusion coefficient,
to the (scaled) subsystem current fluctuation.  Remarkably, the relation
explains why the mass fluctuation in the Manna sandpile must vanish upon
approaching criticality and thus helps characterize the dynamical origin of
hyperuniformity in the system.

\item \textit{Self-diffusion coefficient of tagged particles:}
  We also study the steady-state mean-square displacements of
  tagged-particles, which are characterized by the self-diffusion
  coefficient of the individual particle. We theoretically show that
  the self-diffusion coefficient $\mathcal{D}_s(\rhobar)$ is
  identically equal to the ratio, $a(\rhobar) / \rhobar$, of the
  activity to the global number density of the system, i.e.,
  $\mathcal{D}_s(\rhobar)=a(\rhobar)/\rhobar$, a fluctuation relation,
  which connects the (scaled) tagged-particle displacement fluctuation
  to the density-dependent activity. This relation immediately
  explains a previous simulation observation of Ref. \cite{Cunha_2009}
  that, upon approaching criticality, the self-diffusion coefficient
  in the conserved Manna sandpile vanishes in the same fashion as the
  activity. 
  Our theoretical results are in a reasonably good agreement with
  simulations. 
  
\end{enumerate}

The plan of the paper is as follows. In sec.~\ref{sec:models}, we
define the conserved Manna sandpile and the various quantities of
interest.  In sec.~\ref{sec:theory}, we present our calculation method
where we introduce an approximate truncation scheme, helping us to
calculate the dynamic correlations. Then in sec.~\ref{sec:intj
  fluctuation}, we study the dynamic properties of bond current
fluctuations in the system and provide a scaling argument to explain
its behavior near criticality, followed by the calculation of the
corresponding power spectrum in sec.~\ref{sec:instantaneous current
  correlations}. We then proceed to calculate the variance of the
cumulative subsystem (i.e., space-time integrated) current and
elucidate its relationship with the particle mobility in
sec.~\ref{sec:sptm current fluctuation}. We study the self-diffusion
coefficient of tagged particles and the power spectrum for subsystem
mass fluctuation in sec.~\ref{sec:tagged particle} and
sec.~\ref{sec:mass fluctuation}, respectively. Finally, in
sec.~\ref{sec:conclusion}, we summarize with some concluding remarks.

\section{Dynamic correlations in the steady state}

\subsection{Model and definitions}
\label{sec:models}

We consider the continuous-time variant \cite{Dickman_2001} of the
conserved (``fixed energy'') Manna sandpile~\cite{Manna_1991} on a  
ring of  $L$ sites. Any site $i$, with $i = 0$, $1$, $\ldots$, $L-1$,
can have $m_i \ge 0$ number of particles, with  $m_i = 0$, $1$, $\ldots$,
$N$; the total number of particles 
\begin{align}
  \label{eq:mass conservation}
  N =   \sum_{i=0}^{L-1} m_i,
\end{align}
remains {\it conserved}; in this paper we throughout denote the global
density as $\rhobar = N/L$.
The dynamical rules are as following:
An {\it active} site - a site with $m_i > 1$ -
topples with rate $1$ by randomly, and independently, transferring each of the two particles to one of its nearest neighbours.

The sytem violates detailed balance and eventually reaches
a nonequilibrium (quasi-)steady state, which is not described by the familiar
equilibrium Boltzmann-Gibbs distribution and whose probability measure
 is apriori unknown.   The steady state  of the system is usually 
 charactarized through a global order parameter, called the activity
 $a(\rhobar)$, defined as the density of active sites, 
\begin{align}
  \label{eq:activity def}
  a(\rhobar)= \frac{\angular{N_a}}{L},
\end{align}
where $N_a$ is the total number of active sites in the system and
$\langle . \rangle$ denotes the steady-state average.
Interestingly the system has a
nontrivial spatio-temporal structure and, upon tuning the global density
$\rhobar$, undergoes an absorbing phase transition. That is, above a
critical density $\rho_c$, there are dynamical activities in the
system, but, below the critical density, the dynamical activities
in the steady state cease and consequently there are no movements of particles
in the system. The absorbing phase transition in the conserved Manna
sandpile has been intesively studied in the past and can be charactarized
by the following critical exponents -
the order-parameter exponent $\beta$, the correlation-length exponent
$\nu_\perp$ and the dynamic exponent $z$: Up on approaching criticality
from above, we have the following scaling behavior of
activity $a(\Delta) \sim \Delta^\beta$, correlation
length $\xi \sim \Delta^{-\nu_\perp}$, and the relaxation time $\tau_r \sim L^z$
where relative density $\Delta = \rhobar - \rho_c > 0$ and we use the
critical density $\rho_c \approx 0.94885$, as estimated in
Ref. \cite{Dickman_2001}, throughout our paper.

One can write update rules in an infinitesimal
time-interval between time $t$ and $t+\dd{t}$ as given below,
\begin{align}
  \label{eq:manna update rules}
    m_i(t+\dd{t}) =
  \begin{cases}
    m_i(t) +1 & \frac{1}{2}\hata_{i+1} \dd{t} \\
    m_i(t) +1 & \frac{1}{2}\hata_{i-1} \dd{t} \\
    m_i(t) +2 & \frac{1}{4}\hata_{i+1} \dd{t} \\
    m_i(t) +2 & \frac{1}{4}\hata_{i-1} \dd{t} \\
    m_i(t) -2 & \hata_{i} \dd{t} \\
    m_i(t) & [1-\Sigma \dd{t}],
  \end{cases}
\end{align}
where $\Sigma = (3/4) (\hata_{i+1}+\hata_{i-1}) + \hata_i$; here
$\hata_i$ is an indicator function with
$\hata_i = 1$ if the site is active and $\hata_i = 0$ otherwise. Using
the above  update 
rules eq.\eqref{eq:manna update rules},
we can write the time evolution equation of the first moment of local
mass as 
\begin{align}
  \label{eq:density evolution equation1}
  \dv{t} \angular{m_i(t)} = \left[ \angular{\hata_{i-1}(t) -
  2\hata_{i}(t) + \hata_{i+1}(t) } \right].
\end{align}
Denoting the local density $\rho_i(t) = \angular{m_i(t)}$, we can
alternatively write the above equation as
\begin{align}
  \label{eq:density evolution equation}
  \dv{t} \rho_i(t) =  \sum_{k} \Delta_{i,k} a_k(t)
\end{align}
where $\Delta_{i,k}$ is the discrete Laplacian and
$a_k(t) = \angular{\hata_k}(t)$ is the average instantaneous activity.
On the large  spatio-temporal scales and by taking the diffusive scaling
limit $i \rightarrow x = i/L$ and $t \rightarrow \tau = t/L^2$,
we can write the hydrodynamic time-evolution equation for the local
density field $\rho(x,\tau)$ as in  eq.\eqref{eq:density evolution equation}
\cite{Chatterjee_PRE2018, Tapader_PRE2021},
\begin{align}
  \label{eq:approx density evolution}
  \frac{\partial \rho(x,\tau)}{\partial \tau}
  = \frac{\partial^2 a(\rho)}{\partial x^2}
  \equiv \pdv{x} \qty(D(\rho) \pdv{\rho}{x}),
\end{align}
where $D(\rho)$ is  the density-dependent
bulk-diffusion coefficient. It has been previously demonstrated in Ref.
\cite{Chatterjee_PRE2018, Tapader_PRE2021}, the bulk-diffusion coefficient can be
written in terms of the derivative of the activity $a(\rho)$ wrt density
$\rho$,
\begin{align}
  \label{eq:D and a}
  D(\rho) =  \frac{d a(\rho)}{d \rho} \equiv a^\prime(\rho).
\end{align}
Indeed the relaxation processes occuring on a large (coarse-grained) scale are primarily governed by the bulk-diffusion coefficient - the fact that we later use to introduce a truncation scheme [see eq.\eqref{eq:current approximation}] for analytically calculating various time-dependendent correlation functions, which would not have been possible otherwise in a system with nontrivial correlations  as in sandpiles. Notably, one can recast the density evolution equation~\eqref{eq:density evolution equation} in a microscopic form of the continuity equation,
\begin{align}
  \label{eq:microscopic continuity eq}
  \dv{t} \rho_i(t) = \angular{\mathcal{J}_i(t) -
  \mathcal{J}_{i+1}(t)},
\end{align}
where the microscopic instantaneous current $\mathcal{J}_i(t)$ is
defined as the number of particles crossing a bond $(i,i+1)$ in an
infinitesimal time interval $\qty(t,t+\dd{t})$. It is useful to define a
related observable - the cumulative, or time-integrated, bond current 
$\intc_i(t)$ upto time $t$, which is used to calculate various other
correlation functions, such as that involving mass and activity, and is
easily measured  in simulations.
At the microscopic level, the time-integrated current $\intc_i(t)$
is defined as the total number of particles transferred accross
$i^{th}$ bond, connecting the nearest-neighbour pair of
sites $\qty(i,i+1)$, during a time interval $\qty[0,t]$.
That is, the time-integrated  current accross the $i^{th}$ bond during an 
infinitesimal time interval $\qty[t,t+\dd{t}]$ is simply
$\mathcal{J}_i(t) \dd{t}$ with
\begin{align}
  \label{eq:inst curr def intj}
  \mathcal{J}_i(t) = \lim_{\Delta t \rightarrow 0}
  \frac{\qty(\intc_i(t + \Delta t) - \intc_i(t))}{\Delta t}
  \equiv \frac{d\intc_i(t)}{\dd{t}},
\end{align}
and 
\begin{align}
  \label{eq:total integrated current definition}
  \mathcal{\intc}_i(T) = \int^T_0 \dd{t} \mathcal{J}_i(t).
\end{align}
On the average level, we therefore have 
\begin{align}
  \angular{\mathcal{J}_i(t)} = \angular{\frac{d\intc_i(t)}{\dd{t}}}
  = \frac{d\angular{\intc_i(t)}}{\dd{t}}.
\end{align}
We now decompose the instantaneous current into two parts
- a diffusive component $\mathcal{J}_{i}^{(d)}(t)$ and
a fluctuating component $\mathcal{J}_{i}^{(fl)}(t)$,
\begin{align}
  \label{eq:instantaneous current definition}
  \mathcal{J}_i(t) = \mathcal{J}_{i}^{(d)}(t) +
  \mathcal{J}_{i}^{(fl)}(t), 
\end{align}
where, motivated by eq.\eqref{eq:density evolution equation1}, we identify the
diffusive current as
\begin{equation}
  \label{eq:jdiff def}
  \mathcal{J}_{i}^{(d)}(t) = \hata_i(t) - \hata_{i+1}(t).
\end{equation}
Indeed, as we see later, in that case only the diffusive-current component possesses long-ranged temporal correlations, varying slowly in time as a power law, and the fluctuating component on the other hand is simply a delta-correlated one, explaining the motivation behind the above decomposition.
Note that, due to the fact $\angular{\mathcal{J}_i} =
\angular{\mathcal{J}_{i}^{(d)}} = \angular{\hata_i} - \angular{\hata_{i+1}}$,
we must have $\angular{\mathcal{J}_{i}^{(fl)}} = 0$. Indeed the fluctuating current component can be related to the (conserve) noise term in an appropriately coarse-grained fluctuating hydrodynamic theory, which can be then used to study the large-scale fluctuation properties of the system \cite{MFT_RMP2015, Sadhu_2016}.

In the following sections, we study the fluctuation properties of various compenents of currents, instantaneous and the fluctuating one as decomposed in eq.\ref{eq:jdiff def}. 
Throughout the paper, we use the following notation for correlation function $C_r^{A B}(t,t^\prime)$ involving any two local observables $A_i(t)$ and $B_j(t^{\prime})$ with $t \ge t'$,
\begin{align}
  \label{eq:two point corr}
  C_{r=|i-j|}^{A B}(t,t^\prime) = \angular{A_i(t)
  B_j(t^\prime)} - \angular{A_i(t)} \angular{B_j(t^\prime)},
\end{align}
where $r=|j-i|$ is the relative distance.
Further, we denote the spatial Fourier transform of the correlation function
$C_r^{AB}(t,t^\prime)$ as 
\begin{align}
  \label{eq:first fourier}
  \tilde{C}_q^{AB}(t,t^\prime) = \sum\limits_{r=0}^{L-1}
    C_r^{AB}(t,t^\prime) e^{\vb*{i} q r},
\end{align}
where $q=2\pi k/L$ and $k=0$, $1$, $\ldots$, $L-1$
and the inverse Fourier transform as
\begin{align}
  \label{eq:inverse fourier}
  C_r^{AB}(t,t^\prime) = \frac{1}{L} \sum\limits_{q}
  \tilde{C}_q^{AB} (t,t^\prime) e^{- \vb*{i} q r}.
\end{align}
By introducing a truncation scheme as discussed below, we can theoretically compute the
statistics of different combinations of various local currents $\mathcal{J}_i$, $\mathcal{J}_{i}^{(d)}$,  $\mathcal{J}_{i}^{(fl)}$ as well as mass $m_i$, essentially in terms of the following two correlation  functions - $\angular{\mathcal{Q}_i(t) \mathcal{Q}_j(t^\prime)}$ and $\angular{m_i(t) \mathcal{Q}_j(t^\prime)}$.

\subsection{Theory}
\label{sec:theory}

For the conserved Manna sandpile, we write the stochastic update equation of the integrated current $\intc_i(t)$ in an infinitesimal time (continuous) interval $\qty[t,t+\dd{t}]$,
\begin{align}
  \label{eq:appendix integrated current update equation}
  \intc_i(t+\dd{t}) =
  \begin{cases}
    \textbf{\textit{events}} & \textbf{\textit{probabilities}} \\
    \intc_i(t) + 1 & \frac{1}{2} \hata_i(t) \dd{t} \\
    \intc_i(t) + 2 & \frac{1}{4} \hata_i(t) \dd{t} \\
    \intc_i(t) - 1 & \frac{1}{2} \hata_{i+1}(t) \dd{t} \\
    \intc_i(t) - 2 & \frac{1}{4} \hata_{i+1}(t) \dd{t} \\
    \intc_i(t) & 1-\Sigma \dd{t}
  \end{cases}
\end{align}
where $\Sigma = \qty(\hata_i(t) - \hata_{i+1}(t))$. Using the
above update  rules, the time-evolution equation for the first moment of
the time-integrated current $\intc_i(t)$ can be written as
\begin{align}
  \label{eq:appendix intc evolution eqn}
  \dv{t}\angular{\intc_i(t)} = \angular{\hata_i(t)} -
  \angular{\hata_{i+1}(t)}. 
\end{align}
Similarly, using eq.\eqref{eq:appendix integrated current update equation}, we find the time-evolution equation for the second moment $\angular{\intc_i(t)
\intc_{i+r}(t^\prime)} = C^{\intc\intc}_r(t,t^\prime)$ of the integrated current at two different times $t$ and $t^\prime$, for $t > t^\prime$, as given below (see
appendix~\ref{sec:diff time intc intc corr} for
details),
\begin{align}
  \label{eq:diff time intc intc corr update eqn}
  \dv{t} C^{\intc\intc}_r(t,t^\prime) = \qty(C^{\hata\intc}_r(t,t^\prime) -
  C^{\hata\intc}_{r-1}(t,t^\prime)).
\end{align}
The above equation, which is central to our study, is however difficult to solve exactly due to an infinite hierarchy of correlation functions involved and so we employ below an approximation scheme.

We note that the evolution of space and time dependendent activity-current correlation function $C^{\hata\intc}_r(t,t^\prime)$ appearing in eq.\eqref{eq:diff time intc intc corr update eqn}
contains higher-order correlation functions, involving activity, current and some other observables. The calculations of the higher-order correlations would eventually result in a rapidly growing complexity in the hierarchy of correlation functions, which do not constitute a closed set of equations. More specifically, one can start with the infinitesimal time update equation for $\hata_i$ itself, 
\begin{align}
  \label{eq:activity update eqn}
  \hata_i(t+\dd{t}) =
  \begin{cases}
    \textbf{\textit{events}} & \textbf{\textit{probabilities}} \\
    \hata_i(t) + 1 & \frac{1}{2} \hata_{i+1}(t)
    \hat{p}_i(t) \delta_{m_i,1} \dd{t} \\
    \hata_i(t) + 1 & \frac{1}{2} \hata_{i-1}(t)
    \hat{p}_i(t) \delta_{m_i,1} \dd{t} \\
    \hata_i(t) + 1 & \frac{1}{4} \hata_{i+1}(t)
    \hat{p}_i(t) \dd{t} \\
    \hata_i(t) + 1 & \frac{1}{4} \hata_{i-1}(t)
    \hat{p}_i(t)  \dd{t} \\
    \hata_i(t) - 1 & \hata_i(t) \qty(\delta_{m_i,2} + \delta_{m_i,3})
                     \dd{t} \\ 
    \hata_i(t) & 1-\Sigma \dd{t},
  \end{cases}
\end{align}
where  $\hat{p}_i(t) = \qty(1 - \hata_i(t))$ and
$\Sigma =  \qty(\hat{p}_i(t) / 2) (\delta_{m_i,1} + 1/2)
\qty[\hata_{i+1} + \hata_{i-1}] + \hata_i \qty(\delta_{m_i,2} +
\delta_{m_i,3})$. 
From this update equation, for $t > t^\prime$, we can 
 write the evolution equation for $C^{a \intc}_r(t,t^\prime)$ as
\begin{align}
  \label{eq:a intc update eqn}
  \dv{t} C^{a \intc}_r(t,t^\prime) \equiv \dv{t} \angular{\hata_0(t)
  \intc_r(t^\prime)}_c 
  = \angular{\qty{\dv{t} \hata_0(t)}
  \intc_r(t^\prime)}_c, 
\end{align}
where the observable inside the curly bracket evolves according to the equation,
\begin{align}
  \label{eq:activity evl eqn}
  &\dv{t} \hata_0(t) = - \hata_0(t)
    \qty(\delta_{m_0,2} + \delta_{m_0,3}) +
    \nonumber \\
  &  \qty[\hat{p}_0(t) \qty(\hata_{1}(t) +
    \hata_{L-1}(t))] \qty(\frac{1}{2} \delta_{m_0,1} +\frac{1}{4}).
\end{align}
Clearly, to solve for $C_r^{a \intc}(t,t^\prime)$, one needs to calculate the
correlation functions $\angular{\delta_{m_0,2} \hata_0(t)
  \intc_r(t^\prime)}$ and $\angular{\delta_{m_0,1} \hata_0(t)
  \hata_{1}(t) \intc_r(t^\prime)}$, which would in turn
involve another set of even higher-order correlation functions;
of course, in this way, one generates an infinite hierarchy of equations,
which is difficult to handle.

We bypass the difficulty by employing the following
approximation scheme, which in fact truncates the otherwise infinite
hierarchy. To this end, we approximate the local
diffusive current, which is the gradient of the instantaneous local activity 
and written as
\begin{align}
  \label{eq:current approximation}
  \mathcal{J}_i^{(d)}(t) = \qty[\hata_i(t) - \hata_{i+1}(t)]
  \equiv D(\rhobar) \qty[m_i(t) - m_{i+1}(t)],~~~~
\end{align}
where the bulk-diffusion coefficent is given by $D(\rhobar) = a^\prime(\rhobar)$ as in eq.\eqref{eq:D and a} and can be treated as a constant.
Essentially, we have assumed in the above truncation scheme that
fluctuations around the global density $\rhobar$ are small and the
local current (the gradient of activity) relaxes diffusively via the
gradient of local mass. As demonstrated later, this approximation
captures relevant correlations quite well on the large (hydrodynamic)
time scales.
More specifically, to compute correlations between any observable
$A(t)$ (e.g., current or mass) and  the current observable
$\mathcal{J}_i^{(d)}(t)$, we replace the diffusive-current observable
$\mathcal{J}_i^{(d)}(t)$ by its truncated form  as  given in the rhs of
eq\eqref{eq:current approximation}, to obtain the following equality
(approximate),
\begin{equation}
\langle A(t) \mathcal{J}_i^{(d)}(t') \rangle \simeq 
\left\langle a'(\rhobar) \qty[A(t) \{ m_i(t') - m_{i+1}(t') \}] \right\rangle.
\end{equation}
Now we can proceed further by first rewriting eq.\eqref{eq:diff time intc intc corr update eqn} as
\begin{align}
  \label{eq:diff time intc corr update eqn}
  \dv{t} C^{\intc\intc}_r(t,t^\prime) =
  a^\prime(\rhobar)\qty(C^{m\intc}_r(t,t^\prime) - 
  C^{m\intc}_{r-1}(t,t^\prime)),
\end{align}
and then expressing the time-evolution equation for the mass-current correlations
$C^{m\intc}_r(t,t^\prime) = \angular{m_{i}(t)\intc_{i+r}(t^\prime)} - \langle m_{i}(t) \rangle \langle\intc_{i+r}(t^\prime) \rangle$  as
\begin{align}
  \label{eq:mass current evl eqn}
  \dv{t} C^{m\intc}_r(t,t^\prime) \simeq
  a^\prime(\rhobar) \sum\limits_k \Delta_{r,k}C^{m\intc}_k(t,t^\prime),
\end{align}
where have used eq.\eqref{eq:current approximation} in the intermediate steps
(see appendix sec. \ref{sec:diff time mass intc corr}).
It is worth noting that, in eq.\eqref{eq:current approximation} or in eqs.\eqref{eq:diff time intc corr update eqn} and \eqref{eq:mass current evl eqn}, the activity appears simply as a global density-dependent constant prefactor  $a(\rhobar)$ and thus we obtain a closed set of equations, involving only mass and current correlations; equations \eqref{eq:diff time intc corr update eqn} and \eqref{eq:mass current evl eqn} can be solved, albeit  in terms of the activity $a(\rhobar)$, which however remains undetermined in our theory. Interestingly, it was previously possible to exactly calculate the dynamic correlation functions in  simple exclusion processes \cite{Sadhu_2016} because, in that case, one already gets a closed set of equations for mass and current correlations and, furthermore, because the steady-state measure is a product one, allowing one to calculate various static (time-independent) quantities, which enter into the expression of the current and mass fluctuations. Without an explicit  knowledge of the steady state measure~\cite{Sadhu_JSP2009}, the Manna sandpile, on the other hand, is nontrivial due to the nonzero spatial correlations present in the system, making explicit calculations of the static quantities, such as the density-dependent activity, quite difficult. Nevertheless, as shown below, by extending the formalism developed in the context of simple exclusion processes  \cite{Sadhu_2016}, one can calculate various dynamic correlations in terms of activity and obtain fluctuation relations, which precisely quantify the underlying relationship between dynamic and static fluctuations in the system.

At this stage, it is useful to introduce the Fourier representation of the respective correlation functions as given in eq.\eqref{eq:first fourier}, and we can then write eqs.\eqref{eq:diff time intc corr update eqn} and \eqref{eq:mass current evl eqn} in the respective Fourier modes,
\begin{align}
  \label{eq:basic eqn in fourier1}
  \dv{t} \tilde{C}_q^{\intc \intc} (t,t^\prime) =
  a^\prime (\rhobar) \tilde{C}_q^{m \intc}(t,t^\prime)
  \qty[1-e^{\vb*{i} q}],
\end{align}
and
\begin{align}
  \label{eq:basic eqn in fourier2}
  \dv{t} \tilde{C}_q^{m \intc}(t,t^\prime) =
  -a^\prime(\rhobar) \lambda_q \tilde{C}_q^{m \intc} (t,t^\prime),
\end{align}
where
\begin{equation}
  \label{eq:lamba definition}
  \lambda_q = 2\qty[1-\cos q].
\end{equation}
Now,
eqs.\eqref{eq:basic eqn in fourier1}
and~\eqref{eq:basic eqn in fourier2}, can be integrated to have
\begin{align}
  \label{eq:evolution equation intj corr}
  \tilde{C}_q^{\intc \intc} (t,t^\prime) =
  &
    \int\limits_{t^\prime}^t \dd{t^{\prime \prime}}a^\prime(\rhobar)
    \tilde{C}^{m \intc}_q(t^{\prime \prime},t^\prime)
    \qty[1-e^{\vb*{i} q}] \nonumber \\
    +&\tilde{C}_q^{\intc \intc}
       (t^\prime,t^\prime), 
\end{align}
and
\begin{align}
  \label{eq:mass curr diff time corr sol}
  \tilde{C}^{m\intc}_q(t,t^\prime) = e^{-a^\prime(\rhobar)
  \lambda_q (t-t^\prime)} \tilde{C}^{m\intc}_q(t^\prime,t^\prime).
\end{align}
respectively.
However, to fully solve for the unequal-time correlation functions
$\tilde{C}_q^{\intc \intc} (t,t^\prime)$ and $\tilde{C}^{m\intc}_q(t,t^\prime)$,
we need to calculate their respective equal-time counterparts as well.
First we obtain the  evolution equation for the equal-time mass-current 
correlation function $C^{m\intc}_r(t^\prime,t^\prime)$  and, then writing
$C^{m\intc}_r(t^\prime,t^\prime)$  
in the Fourier space (using eq.\eqref{eq:mass intc corr evl eqn}; see
appendix~\ref{sec:mass intc corr evl eqn} for details), we get
\begin{align}
  \label{eq:evl eqn same time mQ corr}
  \dv{t^\prime} \tilde{C}^{m\intc}_q(t^\prime,t^\prime) =
  -a^\prime(\rhobar) \lambda_q \tilde{C}_q^{m \intc}(t^\prime,t^\prime)
  + \tilde{f}_q(t^\prime),
\end{align}
where the Fourier transform of the source term $\tilde{f}_q(t^\prime)$
in the steady state is given by
\begin{align}
  \label{eq:same time mQ source}
  \tilde{f}_q
  = \tilde{C}^{m\hata}_q \qty(1-e^{-\vb*{i} q}) - 2a(\rhobar)
  \qty(1-e^{-\vb*{i} q} ) \qty[1+\frac{\lambda_q}{4}].
\end{align}
Now eq.\eqref{eq:evl eqn same time mQ corr} can be directly integrated
to obtain
\begin{align}
  \label{eq:same time mass intj corr}
  \tilde{C}^{m\intc}_q(t^\prime,t^\prime) =
  \int\limits_0^{t^\prime} \dd{t^{\prime \prime}} e^{-a^\prime(\rhobar)
  \lambda_q (t^\prime-t^{\prime \prime})} \tilde{f}_{q}(t^{\prime \prime}),
\end{align}
substituting which into
eq.\eqref{eq:mass curr diff time corr sol}, we get the
unequal-time mass-current correlation function  in terms of
$\tilde{f}_q$,
\begin{align}
  \label{eq:final form diff time mass intc corr}
  \tilde{C}^{m\intc}_q(t,t^\prime) = \int\limits_0^{t^\prime}
  \dd{t^{\prime \prime}}
  e^{-a^\prime(\rhobar) 
  \lambda_q (t-t^{\prime \prime})} \tilde{f}_{q}(t^{\prime \prime}).
\end{align}
To calculate the above correlation, we need to calculate the
activity-mass correlation as in eq.\eqref{eq:same time mQ
  source}.
Importantly, as shown below, we can calculate the static (time-independent)
activity-mass correlation function exactly in the steady state
(see appendix~\ref{sec:appendix mass activity correlation}). 
Using the steady-state condition  $d C^{mm}_r(t,t) / dt = 0$, we obtain
\begin{align}
  \label{eq:mass activity evl eqn same time}
  \dv{t} C_r^{mm}(t,t) = 
  & \sum\limits_k 2\angular{m_{0} \Delta_{rk} \hata_{k}}_c + B_{r} = 0,
\end{align}
where $B_r$ is the source term having  the form,
\begin{align}
  B_{r}(\rhobar) = & 7a(\rhobar)\delta_{0,r} -4a(\rhobar)
            \qty(\delta_{0,r+1}
            + \delta_{0,r-1})+ \nonumber \\
          &\frac{a(\rhobar)}{2}\qty(\delta_{0,r+2}
            + \delta_{0,r-2}).
\end{align}
Equation~\eqref{eq:mass activity evl eqn same time} can be solved by
employing a generating function,
\begin{align}
  G\qty(z) = \sum\limits_{r = 0}^{\infty} C^{m\hata}_{r} z^r,
\end{align}
for the equal-time mass-activity correlation;
see appendix.\eqref{sec:appendix mass activity correlation}.
Here we directly provide the solution of the generating function in terms of the static density-dependent activity,
\begin{align}
  \label{eq:mass activity gz}
  G\qty(z) = \frac{3 a(\rhobar)}{2} - \frac{a(\rhobar)}{4} z,
\end{align}
implying the mass-activity static correlation to be
\begin{align}
  C^{m\hata}_r =
  \begin{cases}
    \frac{3a(\rhobar)}{2} & \text{for} \hspace{4pt} r=0, \\
    -\frac{a(\rhobar)}{4} & \text{for} \hspace{4pt} \abs{r}=1, \\
    0 & \text{otherwise},
  \end{cases}
\end{align}
which is in fact exact. Then, by writing Fourier transform of the above equation, we have
\begin{align}
  \label{eq:ma corr fourier}
  \tilde{C}_q^{m\hata} = a(\rhobar)+\frac{a(\rhobar)}{4}\lambda_q,
\end{align}
and substituting the above in eq.\eqref{eq:same time mQ source}, we
straightforwardly obtain
\begin{align}
  \label{eq:final Aij fourier}
  \tilde{f}_q = -a(\rhobar)\qty(1-e^{-\vb*{i} q}) \qty(1+\frac{\lambda_q}{4}).
\end{align}

\subsection{Time-integrated current fluctuation}
\label{sec:intj fluctuation}

\begin{figure}[H]
  \centering
  \includegraphics[scale=0.75]{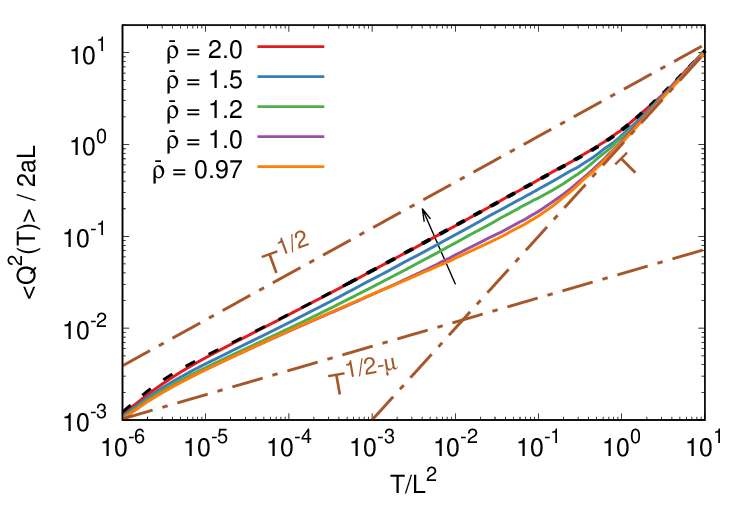}
  
  \caption{Scaled fluctuations of cumulative (time-integrated)
    bond-current up to time $T$, obtained from simulations (solid lines),
    is plotted as a function of scaled time $T/L^2$ for different
    densities $\rhobar = 2.0$ (red), $\rhobar = 1.5$ (blue), $\rhobar =
    1.2$ (green), $\rhobar = 1.0$ (purple), $\rhobar = 0.97$ (orange), and
    for system size $L=1000$, where the arrow
    accross the solid lines denotes the increasing order of
    $\rhobar$; theory as in eq.\eqref{eq:intc corr final} with
    $r=0$, $t=t^\prime =T$ (black dashed line) is in an excellent
    agreement with simulation for $\rhobar = 2.0$. Three guiding
    dot-dashed lines signify the initial-time subdiffusive growth
    $\angular{Q^2(T)} \sim T^{1/2}$ [as in the first part of
    eq.\eqref{eq:integrated current asymptotic}] and the late-time
    diffusive growth $\angular{Q^2(T)} \sim T$ [as in the second part of
    eq.\eqref{eq:integrated current asymptotic}] away from criticality,
    and the initial-time anomalously suppressed subdiffusive growth
    $\angular{Q^2(T)} \sim T^{1/2-\mu}$ [as in eq.\eqref{eq:mu def}] near
    criticality.}
  
  \label{fig:integrated current asymptotics}
\end{figure}

In this section, we calculate the time-integrated bond-current fluctuation, by using the theory developed in the previous section. 
To this end, we substitute eq.\eqref{eq:final Aij fourier} into
eq.\eqref{eq:final form diff time mass intc corr} and get an explicit
solution the first term on the rhs of
eq.\eqref{eq:evolution equation intj corr}. Similarly, we can 
calculate the second term on the rhs of 
eq.\eqref{eq:evolution equation intj corr} as given below,
\begin{align}
  \label{eq:same time curr corr fourier}
  & C^{\intc \intc}_{r}(t,t) \simeq
  \int\limits_0^t \dd{t^\prime}  \Gamma_{r}(t^\prime) + \nonumber \\
  & \frac{1}{L}\int\limits_0^t \dd{t^\prime}
    a^\prime(\rhobar) \sum\limits_{q=0}^{L-1}
  \tilde{C}^{m\intc}_{q}(t^\prime,t^\prime)
  \qty[1 - e^{\vb*{i} q}]\qty(2-\lambda_{qr}),
\end{align}
where $\lambda_{qr} = 2 [1 - \cos(qr)]$;
see appendix sec~\ref{sec:same time intc intc} for details.
Here the quantity $\Gamma_r(t)$ is the strength of the steady-state
correlation function for the fluctuating current
$\angular{\mathcal{J}_0^{(fl)}(t) 
  \mathcal{J}_r^{(fl)}(t^\prime)} = \Gamma_r(t)
\delta(t-t^\prime)$ as derived later in eq.\eqref{eq:fl current corr sol}.
Since we are interested only in the steady-state properties, the 
strength $\Gamma_r$ is replaced by its steady-state value,
\begin{align}
  \label{eq:gammar def}
  \Gamma_r(\rhobar) = 3a(\rhobar) \delta_{0,r} -
  \frac{a(\rhobar)}{2} \qty(\delta_{0,r+1}+\delta_{0,r-1}),
\end{align}
which depends on global density $\rhobar$, through the density-dependent activity $a(\rhobar)$; for the detail calculation of the strength $\Gamma_r$, see 
section~\ref{sec:instantaneous current correlations}. Now the Fourier
transform of  eq.\eqref{eq:same time curr corr fourier}  leads to the second term of rhs of eq.\eqref{eq:evolution equation intj corr}. By using the inverse Fourier transform of eq.\eqref{eq:evolution equation intj corr}, we finally obtain the desired space- and time-dependent current correlation function in the steady-state,
\begin{widetext}
  \begin{align}
    \label{eq:intc corr final}
   C^{\intc\intc}_r(t,t^\prime) =
   & t^\prime \Gamma_r - 
    a^\prime(\rhobar)a(\rhobar)\frac{1}{L}  \sum\limits_{q}
     \int\limits_{0}^{t^\prime} \dd{t^{\prime \prime}}
     \int\limits_{0}^{t^{\prime \prime}} 
     \dd{t^{\prime \prime \prime}}
    e^{-a^\prime(\rhobar)\lambda_q(t^{\prime \prime} - t^{\prime
     \prime \prime})} 
    \lambda_q \qty(1+\frac{\lambda_q}{4}) (2-\lambda_{qr})
     \nonumber \\
    & -a^\prime(\rhobar)a(\rhobar)\frac{1}{L}  \sum\limits_{q}
     \int\limits_{t^\prime}^{t} \dd{t^{\prime \prime}} \int\limits_{0}^{t^\prime}
     \dd{t^{\prime \prime \prime}}
    e^{-a^\prime(\rhobar)\lambda_q(t^{\prime \prime} - t^{\prime
     \prime \prime})} 
      \lambda_q \qty(1+\frac{\lambda_q}{4})
      e^{-\vb*{i} qr}.
\end{align}
\end{widetext}
The asymptotic behavior of the above equation can be
straightforwardly obtained as given below
(see appendix~\ref{sec:intj fluc asymptotic} for details),
\begin{align}
  \label{eq:integrated current asymptotic}
  \angular{\intc^2(T)} \simeq
  \begin{cases}
    \frac{2 a(\rhobar)}{\sqrt{\pi a^\prime(\rhobar)}} T^{\frac{1}{2}}
    & \text{for} \hspace{4pt} 1 \ll T \ll L^2,    \\
    \frac{2a(\rhobar)}{L} T
    & \text{for} \hspace{4pt}T \gg L^2.
  \end{cases}
\end{align}
In simulation, we verify a special case of eq.\eqref{eq:intc corr final} by putting $r = 0$ and $t = t^\prime \equiv T$, i.e., the time-integrated bond current fluctuation $\angular{\intc^2(T)} \equiv C_0^{\intc \intc}(T,T)$ (here average
current $\angular{\intc(T)} = 0$ for the steady-state measurement). In
fig.\ref{fig:integrated current asymptotics}, we plot
$\angular{Q^2(T)}$, obtained from simulation (plotted in solid lines),
for various densities $\rhobar=2.0$ (red), $\rhobar=1.5$ (blue),
$\rhobar=1.2$ (green), $\rhobar=1.0$ (purple), $\rhobar=0.97$
(orange) as a function of $T$. The arrow accross the
solid lines in the fig.\ref{fig:integrated current asymptotics}
signifies the increasing order of global density $\rhobar$.

Indeed the  dynamical behaviors as predicted by the asymptotics in
eq.\eqref{eq:integrated current asymptotic} are different in two
different time regimes: On smaller initial time scales
$1 \ll T \ll L^2$, the time-integrated current grows sub-diffusively as
$T^{1/2}$ and, on larger (hydrodynamic) time scales $T \gg L^2$, grows
linearly as $T$. We  compare the simulation result
with that obtained from our theory eq.\eqref{eq:intc corr final} with $r=0$ and
$t=t^\prime=T$, for $\rhobar=2.0$ (black dashed line) and for system size $L=1000$; one can see an excellent agreement between simulation and theory.

As mentioned previously, our theory is expected to be valid at hydrodynamic times ($T \gg L^z$) and the small-time ($T \ll L^z$) behavior of current fluctuation, especially near criticality, is not quite well captured by eq.{\ref{eq:intc corr final}). However the small-time behavior near criticality can still be obtained qualitatively by using the following standard scaling analysis.
Indeed, first resorting to a simple dimensional analysis, we can see that the activity scales as
$a(\Delta) \sim \Delta^\beta \sim T^{-\beta / \nu_{\perp} z}$, where the relative density $\Delta = \rhobar-\rho_c \ll 1$ and we use the following scaling relations: correlation length $\xi \sim T^{1/z}$ the relative 
density $\Delta \sim \xi^{-1/\nu_\perp} \sim T^{-1/\nu_{\perp} z}$,
with $z$ being the dynamic exponent. Thus, by writing $a(\Delta) / \sqrt{a^\prime(\Delta)} \sim T^{-\mu}$, we straightforwardly have, in the initial-time regime $1 \ll T \ll L^z$, the scaling behavior of the current fluctuation $\angular{\intc^2(T)} \sim ( a/\sqrt{a^\prime}) T^{1/2} \sim T^\alpha$, where the exponents
\begin{align}
  \label{eq:alpha def}
  \alpha = \frac{1}{2} - \mu
\end{align}
and
\begin{align}
  \label{eq:mu def}
  \mu = \frac{\beta + 1}{2 \nu_{\perp} z}.
\end{align}
More precisely,  near criticality we expect the following scaling form for the time-integrated bond-current fluctuation to hold,
\begin{align}
  \label{eq:current scaling ansatz}
  \angular{\intc^2(T)}
  \simeq  L^{\alpha z} \mathcal{G}\qty(\Delta L^{1/\nu_\perp},
  \frac{T}{L^z}) = T^{\alpha}
  \mathcal{F} \qty(\Delta L^{1/\nu_\perp}, \frac{T}{L^z}),
\end{align}
where $\mathcal{G}$ and $\mathcal{F}$ are two scaling functions.
To determine the exponent $\alpha$ in eq.\eqref{eq:current scaling
  ansatz} from simulations, we take density value such that $\Delta
L^{1/\nu_\perp} \rightarrow 0$ and then we plot in
Fig.~\ref{fig:current scaling analysis} the scaled variance of
time-integrated bond current $T^{-\alpha} \angular{\mathcal{Q}^2(T)}$
as a function of the scaled time $T/L^z$ for rather quite smaller
system sizes $L=100$ (solid magenta line), $L=200$ (green dashed line)
and $L=500$ (blue dotted line), respectively and for $\rhobar = 0.95$. We get a reasonably good scaling collapse of simulation data, with the exponent estimated to be $\alpha \approx 0.297$; our theoretical prediction of the exponent $\alpha \approx 0.26$, computed from scaling relations eqs.\eqref{eq:alpha def} and \eqref{eq:mu def} by using $\beta \approx 0.42$, $z \approx 1.66$ and $\nu_\perp \approx 1.81$~\cite{Dickman_2001}, slightly underestimates that obtained from simulation though.

\begin{figure}[H]

  \centering
 \includegraphics[scale=0.75]{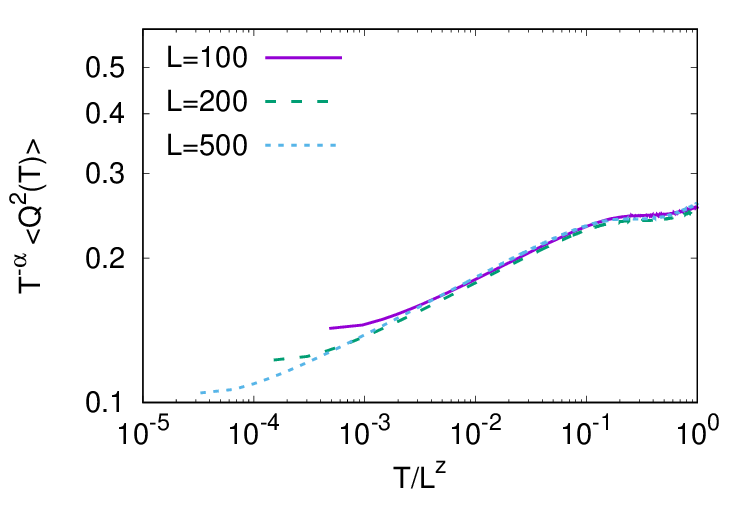}
  \caption{The scaled variance $T^{-\alpha}
    \angular{\mathcal{Q}^2(T)}$ of time-integrated bond current
    $\mathcal{Q}(T)$ up to time $T$, obtained from simulations, is
    plotted as a function of scaled time $T/L^z$ for system size
    $L=100$ (solid magenta line), $L=200$ (green dashed line) and
    $L=500$ (blue dotted line) and  for (near-critical) density
    $\rhobar = 0.95$, where, to achieve the scaling collapse, we use
    $\alpha \approx    0.297$  and $z \approx 1.66$. The value of
    $\alpha \approx 0.297$ obtained from simulations is not far from
    $\alpha \approx 0.26$ obtained from theory as in
    eqs.\eqref{eq:alpha def} and \eqref{eq:mu def} with $\beta \approx
    0.42$, $\nu_\perp \approx 1.81$~\cite{Dickman_2001}. 
  }
  
  \label{fig:current scaling analysis}
\end{figure}

The variance of time-integrated bond current is known to grow
subdiffusively in the initial-time regime in diffusive systems, such as symmetric simple exclusion processes \cite{Masi_2002,Sadhu_2016}. 
Now, away from criticality, the relaxation processes in sandpiles are
diffusive \cite{Chatterjee_PRE2018,Tapader_PRE2021} and, therefore not
surprisingly, the current  fluctuation exhibits a subdiffusive
growth, as derived in the first part of
eq.\eqref{eq:integrated current asymptotic}. Near-critical relaxation processes  in the Manna sandpile, on the other hand, are anomalous and we observe nontrivial scaling behavior. When compared to a normal diffusive system, the Manna sandpile near criticality exhibits strong suppression of current fluctuations and much slower subdiffusive growth of temporal fluctuations due to the lack of local activity, as described in eq.\eqref{eq:current scaling ansatz}; we call it a ``dynamic hyperuniformity'', which is quite analogous to hyperuniformity studied in the spatial domain~\cite{Torquato_2003, Levine_PRL1_2015} and is reminiscent of that identified in the context of temporal statistics of avalanches in a particular version sandpile, called the Oslo ricepile \cite{Garcia-Millan_2018}.
Of course, the dynamic hyperuniformity, or the anomalously subdiffusive growth of temporal fluctuations, can be equivalently characterized in terms of the current and mass power spectra, or the respective dynamic correlation functions, as discussed in the following sections.

\subsection{Current fluctuation and its power spectrum}

\subsubsection{Instantaneous current}
\label{sec:instantaneous current correlations}

In this section, we  calculate in the steady-state the time-dependent
(unequal-time) current-current correlation
function $C_r^{\mathcal{J} \mathcal{J}} (t) \equiv C_r^{\mathcal{J}
  \mathcal{J}} (t,t^\prime=0)$ of the instantaneous  bond current by taking
time derivative of time-integrated bond current correlation as given below,
\begin{align}
  \label{eq:inst current correlation}
  C^{\mathcal{J} \mathcal{J}}_{r}(t) = 
  \qty[\dv{t} \dv{t^\prime} C^{\intc\intc}_r(t,t^\prime)]_{t^\prime=0},
\end{align}
where $t \geq t^\prime$. Now, after differentiating eq.\eqref{eq:intc corr
  final}, 
we can write the time-dependent current correlation as
\begin{align}
  \label{eq:t>s curr corr}
  & C^{\mathcal{J} \mathcal{J}}_{r}(t)  =
    \Gamma_r \delta(t) - \nonumber \\
  &a^\prime(\rhobar)a(\rhobar)\qty[\frac{1}{L}
  \sum\limits_{q}
  e^{-a^\prime(\rhobar)\lambda_q t} 
    \lambda_q \qty(1+\frac{\lambda_q}{4}) 
    e^{-\vb*{i} qr}],
\end{align}
where $\Gamma_r$ is the strength of the fluctuating current
$\mathcal{J}^{fl}(t)$ and calculated below
(see eq.\eqref{eq:fl current corr sol}).
We note that, as $\lambda_q \geq 0$ for any $q$, the  current
correlation function is \textit{negative} $C^{\mathcal{J}
\mathcal{J}}_{r}(t) < 0$ for any $t > 0$. Moreover, provided that we first
take the infinite-system-size limit $L \rightarrow \infty$, the
time-integrated bond current correlation function $C^{\mathcal{J}
\mathcal{J}}_{0}(t)$ over a large time interval $[-T,T]$ decays as
a function of time $T$ as given below,
\begin{align}
  \label{eq:integrated current correlation}
  \int\limits_{-T}^{T} C^{\mathcal{J} \mathcal{J}}_{0}(t) \dd{t}
  \simeq
  \frac{a(\rhobar) }{ \sqrt{\pi a^\prime(\rhobar)}} T^{-\frac{1}{2}}.
\end{align}
Finally, by taking the limit $T \rightarrow \infty$, we obtain the
following identity,
\begin{align}
  \label{eq:infinte sptm vol cur corr}
  \int\limits_{-\infty}^{\infty} C^{\mathcal{J} \mathcal{J}}_{0}(t)
  \dd{t} = 0; 
\end{align}
see appendix~\ref{sec:inst curr fluc asymptotics} for details.
Indeed, the above result is a direct consequence of the
negative current 
correlation present in the system and explains why
the time-integrated bond current fluctuation, as derived in 
eq.\eqref{eq:integrated current asymptotic}, grows subdiffusively in
the initial-time regime,
$1 \ll t \ll L^2$. The asymptotic form of the time-dependent
instantaneous current correlation function, for
$t>0$, in the thermodynamic limit can be written as 
\begin{align}
  \label{eq:inst curr asymptotic}
  C^{\mathcal{J}\mathcal{J}}_0(t) \simeq
  -\frac{a(\rhobar)}{4\sqrt{\pi a^\prime(\rhobar)}} t^{-\frac{3}{2}},
\end{align}
where the density dependendent term
$a(\rhobar)/\sqrt{a^\prime(\rhobar)}$ in the prefactor is the same as
that in eq.\eqref{eq:integrated current asymptotic} (see
appendix~\ref{sec:inst curr fluc asymptotics}). Again, by employing the
 dimensional scaling argument used in the previous section to derive eq.\eqref{eq:mu def},  we obtain a modified power-law decay of the instantaneous current correlation near criticality,
\begin{align}
  \label{eq:inst curr asymptotic criticality}
  C^{\mathcal{J}\mathcal{J}}_0(t) \sim
  -t^{-\qty(\frac{3}{2} + \mu)}.
\end{align}
Clearly the decay is faster than that away from criticality.
As discussed previously, the faster decay of the near-critical current
correlation function is due to the fact that the activity is very small 
in the vicinity of criticality, thus resulting in the anomalous suppression 
of fluctuations. Indeed the suppressed fluctuation
is charactarized by  the exponent $\mu > 0$, whereas $\mu=0$ signifies
the subdiffusive growth of the time-integrated current, expected in a normal diffusive systems.

\subsubsection{Fluctuating current}
\label{sec:fluctuating current correlations}

Now we discuss the dynamic properties of the fluctuating part
$\jfl_i(t)$ in the 
instantaneous bond current, which has already been defined 
in eq.\eqref{eq:instantaneous current definition} and whose strength
appears in the actual current correlation functions
(e.g., see eqs.\eqref{eq:integrated current correlation} and
~\eqref{eq:inst current correlation}).
Here we derive the the general space and time dependence of the
correlation function
$C^{\jfl \jfl}_r(t,t^\prime)$ of the fluctuating current
$\jfl_i(t)$, by using the relation [obtained from the
definition in eq.\eqref{eq:instantaneous current definition}],
\begin{align}
  \label{eq:fl current correlation}
  C^{\jfl \jfl}_r(t,0) =
  & C^{\mathcal{J} \mathcal{J}}_r(t,0) -
    C^{\mathcal{J} \jd}_r(t,0) -
    C^{\jd \mathcal{J}}_r(t,0) \nonumber \\
  & + C^{\jd \jd}_r(t,0),
\end{align}
and a second relation
\begin{align}
  \label{eq:fl current proof1}
  C^{\mathcal{J} \jd}_r(t,0) \equiv
  \dv{t} C^{\mathcal{Q} \jd}_r(t,0) =
  C^{\jd \jd}_r(t,0).
\end{align}
We see that the second and the fourth terms of
eq.\eqref{eq:fl current correlation} cancel each
other. Again, by using the following relation, for
$t >t^\prime$, 
\begin{align}
  \label{eq:fl current proof2}
  C^{\jd \mathcal{J}}_r(t,t^\prime)
  = \dv{t^\prime} C^{\jd \mathcal{Q}}_r(t,t^\prime) =\dv{t^\prime}
  \dv{t} C^{\mathcal{Q} \mathcal{Q}}_r(t,t^\prime), 
\end{align} in eq.\eqref{eq:fl current correlation} along with
$t^\prime = 0$, we finally have the time-dependendent correlation
fuction for the fluctuating current,
\begin{align}
  \label{eq:fl current corr sol}
  C^{\mathcal{J}^{\qty(fl)} \mathcal{J}^{\qty(fl)}}_r(t,t^\prime=0)
  \equiv C^{\mathcal{J}^{\qty(fl)} \mathcal{J}^{\qty(fl)}}_r(t) =  
  \delta(t) \Gamma_r(\rhobar),
\end{align}
where $\Gamma_r(\rhobar)$ is the density-dependendent strength of the
fluctuating current $\jfl$.

\begin{figure}[H]

\subfloat[\label{fig:excess current plot}]{\includegraphics[width=0.5\textwidth]{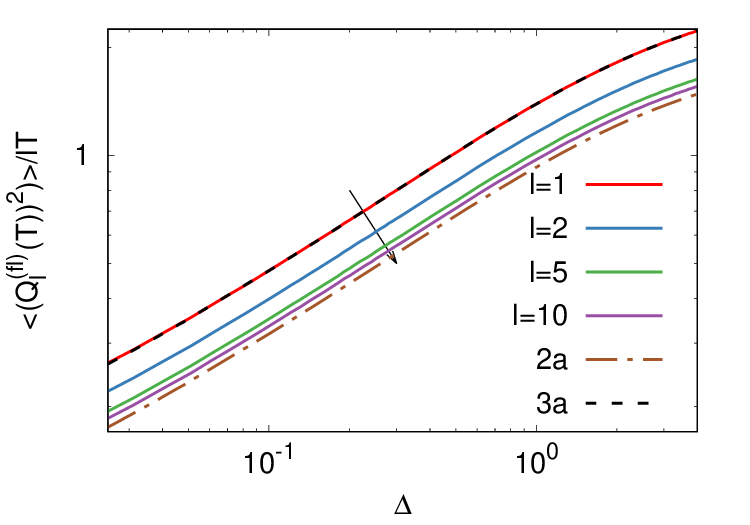}} \\
\subfloat[\label{fig:scaled excess current}]{\includegraphics[width=0.5\textwidth]{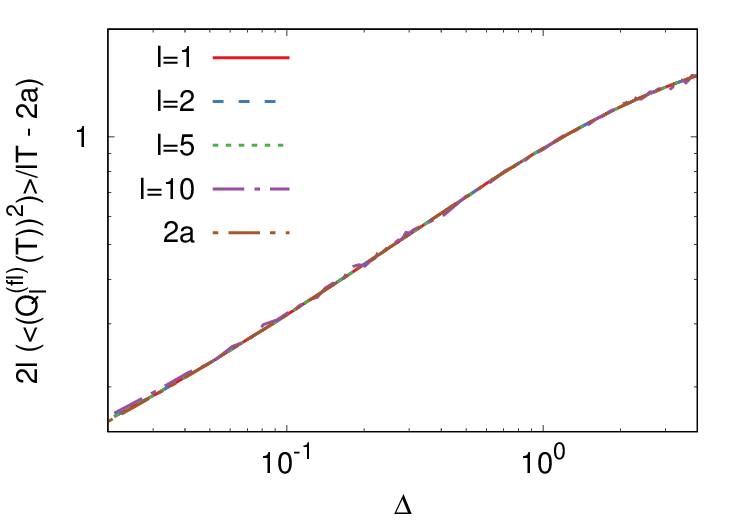}}
  
\caption{\textit{Top panel:} The scaled variance $\langle
  [\mathcal{Q}^{(fl)}_{l}(T)]^2 \rangle /lT$ of
  space-time-integrated fluctuating current
  $\mathcal{Q}^{(fl)}_{l}(T)$ up to time $T$ [as defined  in
  eq.\eqref{eq:fl current numerical verification}], obtained from
  simulations, is plotted as a function of relative density $\Delta$
  for different subsystem sizes $l=1$ (solid red line), $l=2$ (solid
  blue line),  $l=5$ (solid green line), $l=10$ (solid purple line)
  and the arrow accross the solid line
  denotes the increasing order of the subsystem size $l$; we
  have used system size $L=1000$ and final time $T=1000$. The
  theoretical prediction as in 
  eq.\eqref{eq:subsystem fl current variance} with $l=1$ (black dashed
  line) is in an excellent agreement with the corresponding
  simulation; also, one can see that the variance for large $l$
  converges quite rapidly to the theoretically predicted value
  $2a(\rhobar)$ [i.e., eq.\eqref{eq:subsystem fl current variance} for
  $l \gg 1$] (brown dot-dashed line). 
  \textit{Bottom  panel:} A scaling collapse of the scaled variance of space-time-integrated fluctuating current minus the asymptotic value $2 a(\rhobar)$ for different subsystem sizes  is observed, when plotted as a function of relative density $\Delta$, and it is in excellent agreement with theory in eq.\eqref{eq:subsystem fl current  variance}.} 
  \label{fig:excess current single bond}
\end{figure}

The analytical expression of the strength $\Gamma_r$ as given in
eq.\eqref{eq:gammar def} has some interesting properties, which are
due to the two-particle transfer rule in the Manna sandpile and are noticeably different from that in the variant of sandpile with one-particle transfer~\cite{Jain_2005} and symmetric simple exclusion processes studied in Ref.~\cite{Sadhu_2016}. As in the simple exclusion processes, the strength $\Gamma_r$ for the sandpile with one-particle transfer rule can be shown to be simply delta correlated in
space, i.e., $\Gamma_r = 2a(\rhobar) \delta_{r,0}$~\cite{anirbanthesis}; this is because both the models have a steady-state with a product measure
and therefore do not have any spatial correlations.
But, in the case of the Manna sandpile, there are nonzero spatial
correlations, leading to the spatially correlated fluctuating current,
i.e., $\Gamma_r \neq 0$ for $r \neq 0$, as shown in
eq.\eqref{eq:gammar def}. Indeed, as our calculation shows (for
details, see appendix sec.~\ref{sec:same time intc intc}),
eq.\eqref{eq:gammar def} is exact in the case of Manna sandpile and 
we have
\begin{equation}
  \label{eq:gamma properties}
  \Gamma_0 = 3a(\rhobar),
\end{equation}
being the strength of the fluctuating current $\jfl_i$ accross a
single bond $(i,i+1)$.
Moreover, we find that there exsists a sum-rule
\begin{equation}
  \label{eq:gamma r sum}
  \sum\limits_r \Gamma_r = 2a(\rhobar),
\end{equation}
which, as shown later in eq.\eqref{eq:full system intc variance},
is directly related to the scaled space-time integrated current
fluctuations and therefore related to another transport
coefficient, called the mobility, or equivalently, the conductivity, defined
as the ratio between average current and an externally applied small biasing
force~\cite{MFT_RMP2015, Chatterjee_PRE2018}. It has been derived in
Ref.\cite{Chatterjee_PRE2018} that the conductivity in the Manna sandpile is nothing but the density-dependent activity $a(\rhobar)$ itself. Remarkably, as shown in the next section, here we show that one can indeed  relate the conductivity
directly to the current fluctuation in the system.

To verify eq.\eqref{eq:fl current corr sol} in simulation, let us first
define a cumulative (space-time integrated) fluctuating current
across a subsystem of size $l$ and up to time $T$, 
\begin{align}
  \label{eq:fl current numerical verification}
  \mathcal{Q}_{l}^{(fl)}(T) = \int\limits_0^T \dd{t}
  \sum\limits_{i=0}^{l -1}\mathcal{J}_{i}^{(fl)}(t).
\end{align}
Then, using eqs. \eqref{eq:gammar def} and\eqref{eq:fl current corr sol} and after some algebraic manipulations, we obtain, for $l < L$, a fluctuation  relation, which immediately connects the scaled current fluctuation and the density-dependent activity,
\begin{align}
  \label{eq:subsystem fl current variance}
  \frac{1}{l T} \angular{\qty(\intc_{l}^{\qty(fl)}(T))^2}
  =  2a(\rhobar) \qty(1+\frac{1}{2l}).
\end{align}
In fig.\ref{fig:excess current plot}, we
plot the lhs of eq.\eqref{eq:subsystem fl current variance} as a function
of the relative density $\Delta = \rhobar - \rho_c$ for different subsystem
sizes $l=1$ (solid red line), $l=2$ (solid blue line), $l=5$ (solid
green line), $l=10$ (solid purple line); we take system size $L=1000$ and
final 
time $T=100$. The arrow accross the solid lines
  denotes an increasing order of the subsystem size $l$. Note that
the variance of subsystem fluctuating current 
for subsystem size $l=1$ is actually the strength $\Gamma_0$ of the
fluctuating 
bond current; the corresponding analytical result $\Gamma_0 =
3a(\rhobar)$ (dashed black line) as in eq.\eqref{eq:gamma properties}
shows an excellent agreement with simulations. For comparison, in the
same fig.\ref{fig:excess current plot}, we also plot $2a(\Delta)$  as
a function of $\Delta$ (the brown dot-dashed line), to demonstrate
that, as subsystem size $l$ increases, the scaled variance (lhs of
eq.\eqref{eq:subsystem fl current variance}) indeed converges towards
$2a(\rhobar)$, as predicted in eq.\eqref{eq:subsystem fl current
  variance}. To show this convergence more quantitatively, in
fig.~\ref{fig:scaled excess current} we plot the scaled quantity
$2l\qty[\angular{\qty(\mathcal{Q}_l^{(fl)})^2}-2a]$ for various
subsystem sizes $l = 1$ (red solid line), $l=2$ (blue dashed line),
$l=5$ (green dotted line) and $l = 10$ (purple dot-dashed line); we see that all the curves collapse excellently onto each other and the collasped master curve match excellently with the analytically predicted value $2a(\Delta)$ derived in eq.\eqref{eq:subsystem fl current variance}.

\subsubsection{Space-time integrated current}
\label{sec:sptm current fluctuation}

In this section we calculate the steady-state variance $\angular{\bar{Q}^2(l,T)} - \angular{\bar{Q}(l,T)}^2$ of the cumulative (space-time integrated) actual particle current $\bar{Q}(l,T) = \sum_{i=0}^{l-1} \mathcal{Q}_i(T)$ across a subsystem of size $l$ and up to time $T$, which can be written as 
\begin{align}
  \label{eq:sptm intj fluctuations}
  &\angular{\bar{Q}^2(l,T)} - \angular{\bar{Q}(l,T)}^2 =
    \angular{\bar{Q}^2(l,T)} \nonumber \\
  &= lC^{\intc\intc}_0(T,T) + \sum\limits_{r=1}^{l-1} 2(l-r)
  C^{\intc\intc}_r(T,T),
\end{align}
where we have used the fact that the average steady-state current is zero, i.e., $\angular{\bar{Q}(l,T)} = 0$. Now, by using the following identity, 
\begin{align}
  \label{eq:simplification}
  \sum\limits_{r=1}^{l-1} 2(l-r)(2-\lambda_{rn})
  = 2\qty(\frac{\lambda_{ln}-l\lambda_n}{\lambda_n}),
\end{align}
we can rewrite
eq.\eqref{eq:sptm intj fluctuations} as 
\begin{widetext}
  \begin{align}
  \label{eq:final sptm QiQj fluctuations}
  \angular{\bar{Q}(l,T)^2}
  =
  2a(\rhobar)lT + a(\rhobar)T\qty(1-\delta_{l,L}) -
    2a(\rhobar) \frac{a^\prime(\rhobar)}{L}
    \sum\limits_{q}
    \frac{a^\prime(\rhobar)\lambda_q T -1+\exp\qty(-\lambda_q
    a^\prime(\rhobar)T)}{\qty(\lambda_q a^\prime(\rhobar))^2} 
    \lambda_{q}\qty(1+\frac{\lambda_q}{4})
    \frac{\lambda_{ql}}{\lambda_q}. 
\end{align}
\end{widetext}
In fig.\ref{fig:sptm integrated current}, we plot the subsystem current fluctuation
$\angular{\bar{Q}^2(l,T)}$ obtained from simulations as a function of relative
density $\Delta = \rho - \rho_c$ for various subsystem sizes $l$ and
final times $T$: $l=2500$, $T=100$ (upper solid sky-blue line) and
$l=100$, $T=10^5$ (lower solid magenta line). In the same figure we also compare the simulation results with theory eq.\eqref{eq:final sptm QiQj fluctuations}: $l=2500$,
$T=100$ and $l=100$, $T=10^5$ (both in black dashed line); we
observe excellent agreement between simulations and theory. Here we
note that the results for the larger subsystem size $l$ and smaller
$T$ (upper solid line) almost coincide with twice of local activity,
$2a(\Delta)$, as a 
function of $\Delta$ (red dot-dashed line).

Importantly, the asymptotic expression of the variance of cumulative subsystem (space-time
integrated) current as in eq.\eqref{eq:final sptm QiQj fluctuations} depends on the order of limits of the two variables $T \gg 1$ and $l \gg 1$, i.e.,
\begin{align}
  \label{eq:sptm large time fluc}
  \frac{\angular{\bar{Q}^2(l,T)}}{lT} \simeq
  \begin{cases}
    \frac{2a(\rhobar)}{\sqrt{\pi a^\prime(\rhobar)}}
    \frac{l}{\sqrt{T}} & \text{for} \hspace{4pt} T \gg 1, l \gg 1, \\
    2a(\rhobar) - \frac{8a(\rhobar)
    \sqrt{a^\prime(\rhobar)}}{3\sqrt{\pi}} \frac{\sqrt{T}}{l}
                                    & \text{for} \hspace{4pt} l \gg 1,
                                      T \gg 1; 
  \end{cases}
\end{align}
see appendix sec.~\ref{sec:sptm integrated current} for details.
The first expression in the above equation have been obtained by
taking the limit in the following order,
first $T \gg 1$  and then the limit $l \gg 1$. In this particular
order of limits, the scaled fluctuation $\angular{\bar{Q}^2(l,T)} /
lT$ decreases as $1/\sqrt{T}$ and eventually vanishes in the limit of
$T \rightarrow \infty$. On the other hand, if we take the limit in the
opposite order, $l \gg 1$ first and then $T \gg 1$, we obtain the
second asymptotic expression in eq.\eqref{eq:sptm large time fluc}. That is, in the limit $l \rightarrow \infty$, the scaled subsystem-current fluctuation $\angular{\bar{Q}^2(l,T)} / lT$ tends to $2a(\rhobar)$ as one increases $T$,
\begin{align}
  \label{eq:sptm intjc limit}
  \sigma_Q^2 (\rhobar) \equiv \lim\limits_{l \rightarrow \infty}  \lim\limits_{T\rightarrow \infty} 
  \frac{\angular{\bar{Q}^2(l,T)}}{lT} = 2a(\rhobar);
\end{align}
here the infinite-subsystem-size limit is taken first, and then the infinite-time limit.
Note that, in all the above cases, we have taken the large system size limit $L/l \gg 1$ at the very beginning. In fact, one can immediately identify the rhs of eq.\eqref{eq:sptm intjc limit} as the mobility (equivalently, the conductivity) for the Manna sandpile, as calculated in Ref.\cite{Chatterjee_PRE2018}. Indeed, eq.~\eqref{eq:sptm intjc limit} can be thought of as a nonequilibrium version of the celebrated Green-Kubo relations well known for equilibrium systems~\cite{Marconi_2008}.

\begin{figure}[H]
  
  \includegraphics[scale=0.75]{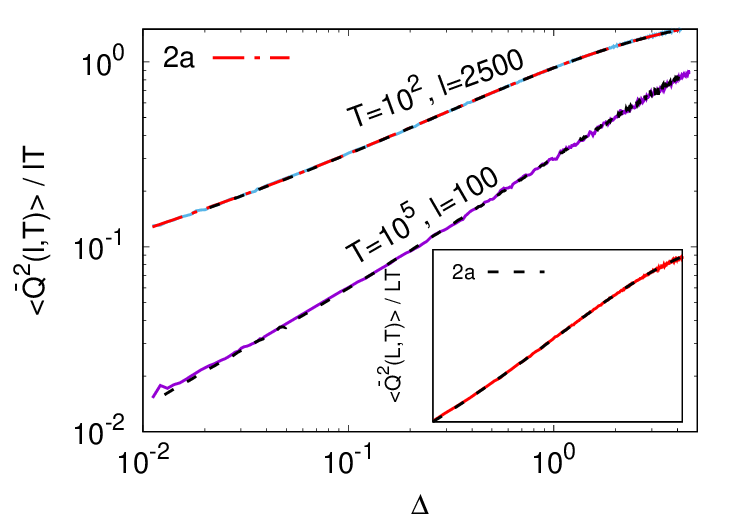}

  \caption{Scaled space-time-integrated current fluctuations as a function of
    relative density. The simulation data for subsystem size $l=2500$
    and $T=100$ is plotted on top in \textit{solid sky-blue line} and data for
    $l=100$ and $T=10^5$ is plotted at bottom in \textit{solid magenta line}. We
    compared the analytical result eq.\eqref{eq:final sptm QiQj
      fluctuations} (corresponding dashed black
    lines) with the simulation, which is in excellent agreement. Simulation
    data have been taken for $L=5000$ in both cases. We note that
    the result for the larger subsystem size $l$ and smaller
    $T$ (upper solid line) almost coincide with twice of local activity,
    $2a(\Delta)$, as a function of $\Delta$ (red dot-dashed line).
    \textit{In the inset} we
    compared the scaled total current fluctuation and twice of
    activity $2a(\rhobar)$ as a function of $\Delta$.}

\label{fig:sptm integrated current}
\end{figure}

Interestingly, if we take $l=L \gg 1$, which corresponds to the bond currents summed over the
whole system, we have the following identity
\begin{align}
  \label{eq:full system intc variance}
  \lim\limits_{L \rightarrow \infty}
  \frac{\angular{\bar{Q}^2(L,T)}}{LT} = 2a(\rho) = \sum\limits_r 
  \Gamma_r.
\end{align}
Notably the above equality is valid for any finite time $T$. This is
because the sum of the diffusive currents 
over the full system, $\sum_{i=1}^L \mathcal{J}_i^{(d)}$, is 
zero by definition (see eq.\eqref{eq:jdiff def}). Consequently the
rhs of eq.\eqref{eq:full system intc variance} is equal to the
space-time inetgral of the fluctuating-current correlation function
$\sum_{r=-\infty}^{\infty} \int_{-\infty}^{\infty}
\dd{t}C^{\mathcal{J}^{\qty(fl)} \mathcal{J}^{\qty(fl)}}_r(t,0) = 2a(\rhobar)$,
obtained using eqs.\eqref{eq:gammar def},~\eqref{eq:fl current corr
  sol} and~\eqref{eq:gamma r sum}. 
In the inset of fig.\ref{fig:sptm integrated current},
the scaled variance $\angular{\bar{Q}^2(L,T)}/LT$ (solid red line) and
twice the local activity $2a(\Delta)$ (black dashed line) are plotted as a
function of $\Delta$, which is in excellent agreement with
eq.\eqref{eq:full system intc variance}.
Clearly, below the critical point $\Delta < 0$, the system goes into
an absorbing state and, as a result, the current fluctuation is identically
zero. Usually activity is considered to be the order parameter in the
sandpiles. Indeed, as the identity eq.\eqref{eq:full system intc variance} suggests, the space-time integrated  current fluctuation can serve as an order  parameter and thus characterizes the dynamical state of the system. Later we show that the self-diffusion coefficient of tagged particles can  be expressed in terms of the activity and, as previously noted in~\cite{Cunha_2009}, it can be considered an alternative description of the system's order parameter.

\subsubsection{Power spectrum}
\label{sec:power spectrum current}

The two-point time-dependent correlation function for instantaneous 
bond current can be characterized also
through the power spectrum analysis, which we perform in this section.
From the \textit{Wiener-Khinchin} theorem~\cite{MacDonald2006}, the
power spectrum for the instantaneous bond current $\mathcal{J}_i(t)$
is expressed in terms of the Fourier transform of the time-correlation functions, 
\begin{align}
  \label{eq:ps wiener khinchin}
  S_{\mathcal{J}}(f) =   \int\limits_{-\infty}^\infty \dd{t} 
  C^{\mathcal{J} \mathcal{J}}_0(t,0) e^{2\pi \vb*{i} f t}.
\end{align}
Setting $r=0$ in  eq.\eqref{eq:t>s curr corr}, we perform the integration in the
rhs of the above equation, leading to the following expression, 
\begin{equation}
  \label{eq:ps simplified form}
  S_{\mathcal{J}}(f) = \frac{2a(\rhobar)}{L} +
  \frac{2a(\rhobar)}{L} 
  \sum_{q} \qty(1+\frac{\lambda_q}{4})
  \frac{4\pi^2f^2}{\lambda_q^2 a^\prime(\rhobar)^2 +
  4\pi^2f^2}.
\end{equation}
Now, by subtracting the $f=0$ mode,
\begin{align}
  \label{eq:ps and current}
  S_{\mathcal{J}}(0) = \lim\limits_{T
  \rightarrow \infty} \frac{\angular{\mathcal{Q}^2(T)}}{T} =
  \frac{2a(\rhobar)}{L},
\end{align}
from the lhs of  eq.\eqref{eq:ps simplified form}, we rewrite eq. \eqref{eq:ps simplified form} in terms of the modified power spectrum $\tilde{S}_{\mathcal{J}}(f) =
S_{\mathcal{J}}(f) - S_{\mathcal{J}}(0)$,
\begin{align}
  \label{eq:power spectrum substracted}
  \tilde{S}_{\mathcal{J}}(f) = \frac{2a(\rhobar)}{L} 
  \sum_{q} \qty(1+\frac{\lambda_q}{4})
  \frac{4\pi^2f^2}{\lambda_q^2 a^\prime(\rhobar)^2 +
  4\pi^2f^2}.
\end{align}
We can now straightforwardly obtain the asymptotic form of eq.\eqref{eq:power spectrum
  substracted} for small frequency $1/L^2 \ll f \ll 1$. To do this, we first replace the sum in eq.\eqref{eq:power spectrum substracted} as an integral over the variable $x=q / 2\pi$,
\begin{align}
  \label{eq:curr ps small}
  \tilde{S}_{\mathcal{J}}(f) \simeq
  4a(\rhobar) 
    \int\limits_{1/L}^{1/2} \dd{x} 
    \frac{1+\frac{\lambda(x)}{4}}{1 +
  \frac{\lambda^2(x) {a^\prime}^2(\rhobar)}{4\pi^2f^2}},
\end{align}
where $\lambda(x) \simeq 4\pi^2x^2$.
Then, by performing the variable transformation
\begin{align}
  \label{eq:dimensionless y}
  y = \frac{\lambda^2(x) {a^\prime}^2(\rhobar)}{4\pi^2f^2},
\end{align}
and doing some algebraic manipulations, we immediately obtain the modified power spectrum of current,
\begin{align}
  \label{eq:ps curr asymptotic}
  \tilde{S}_{\mathcal{J}}(f) \simeq
  a(\rhobar) \sqrt{\frac{f}{2\pi a^\prime(\rhobar)}}
    \int\limits_0^\infty \dd{y} \frac{y^{-\frac{3}{4}}}{(1+y)} =
  \frac{\sqrt{\pi} a(\rhobar)}{\sqrt{a^\prime(\rhobar)}}
  f^{\frac{1}{2}};
\end{align}
see appendix sec.~\ref{sec:identities and integrals} for details.
Again, by using the previous dimensional scaling argument where $a/\sqrt{a^\prime} \sim f^\mu$ with $\mu$ given in eq.\eqref{eq:mu def} (see sec.\ref{sec:intj fluctuation}), we obtain the desired scaling behavior of the subtracted power spectrum near criticality,
\begin{align}
  \label{eq:final largef ps curr near criticality}
  \tilde{S}_{\mathcal{J}}(f) \sim
  f^{\psi_{\mathcal{J}}},
\end{align}
where $\psi_{\mathcal{J}} = 1/2 + \mu$. Since $\mu > 0$, with decreasing frequency, the near-critical power spectrum in the above equation \eqref{eq:final largef ps curr near criticality} decays faster than that away from criticality (given by eq.\eqref{eq:ps curr asymptotic}). In simulations, we calculate the power spectrum by discretizing time over a  small interval $\delta t$ and calculate the discrete Fourier transform 
\begin{align}
  \label{eq:dft current}
  \tilde{\mathcal{J}}_{n;T} = \delta t \sum\limits_{k=0}^{T-1}
  e^{\vb*{i} 2\pi f_n k} \mathcal{J}_i(k),
\end{align}
where $f_n = n/T$ with $T$ being large. Then we define the power spectrum of the bond current as
\begin{align}
  \label{eq:ps current}
  S_{n} = \lim\limits_{T \rightarrow \infty} \frac{1}{T}
  \angular{\abs{\tilde{\mathcal{J}}_{n;T}}^2}. 
\end{align}
In fig.\eqref{fig:current ps}, we plot the substracted power spectrum
$\tilde{S}_{\mathcal{J}}(f)$, obtained from  simulations in solid lines, for various densities
$\rho=2.0$ (red line), $\rho=1.5$ (blue line), $\rho=1.5$ (green
line), $\rho=1.0$ (purple line), $\rho=0.97$ (orange
line). The arrow through the solid lines denotes the
incremental order of the density $\rhobar$. For $\rho=2.0$, we also
plot $\tilde{S}_{\mathcal{J}}(f)$ obtained from theory eq.\eqref{eq:power spectrum substracted} (black dashed line), which shows an excellent agreement with simulation; top-most guiding line - $f^{1/2}$ [behavior away from criticality as in eq.\eqref{eq:ps curr asymptotic}] and the bottom-most guiding line - $f^{1/2 + \mu}$ [behavior near criticality as in eq.\eqref{eq:final largef ps curr near criticality}].

\begin{figure}[H]
  \centering
    \includegraphics[width=0.5\textwidth]{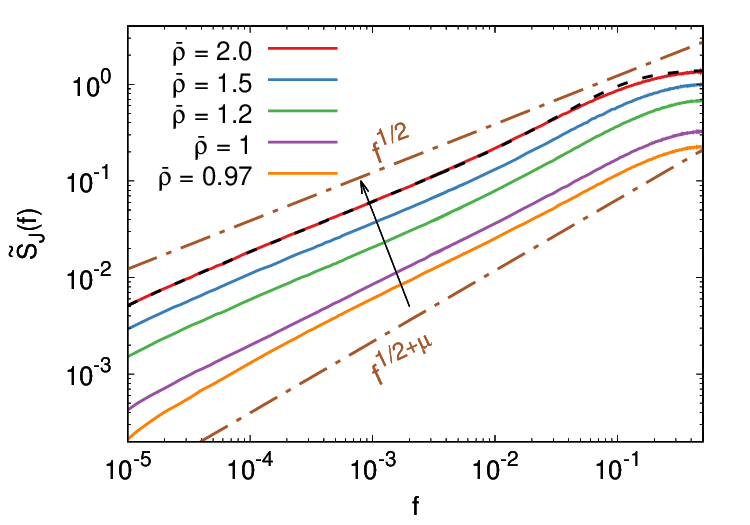}
    \caption{ Power spectrum of instantaneous current, computed from
      simulations is plotted in solid lines as a function of frequency $f$
      for various densities $\rho=2.0$ (red line), $\rho=1.5$ (blue line),
      $\rho=1.5$ (green line), $\rho=1.0$ (purple line), $\rho=0.97$ (orange
      line) and for system size $L=1000$. The arrow accross the solid lines
      signifies the incremental order of the density $\rhobar$. The top and
      bottom dashed guiding lines represent the asymptotic behaviors, the
      $f^{1/2}$ (far from criticality) and $f^{1/2+\mu}$ (near criticality)
      scaling, respectively, with $\mu \approx 0.24$, obtained by using
      $\beta \approx 0.42$, $\nu_{\perp} \approx 1.81$ and $z \approx 1.66$
      in eq. \eqref{eq:mu def}. The dashed black line represents the
      theoretical result eq.\eqref{eq:ps simplified form} for $\rho=2.0$ and
      is in excellent agreement with the correponding simulation (top red
      solid line).  }
    
  \label{fig:current ps}
\end{figure}

\subsection{Tagged-particle displacement fluctuation}
\label{sec:tagged particle}

In this section, we study the fluctuations in tagged particle
displacements as a function of time.
We can relate the sum of all individual time-integrated tagged particle
displacement $\sum_{\alpha=1}^N X_\alpha(T)$, where the net
displacement $X_\alpha(T)$ of $\alpha^{th}$
particle in a time interval $[0,T]$, to the space-time integrated current by
the following relation,
\begin{equation}
  \label{eq:tagged particle intj}
  \sum\limits_{\alpha=1}^N X_\alpha(T) = \sum\limits_{i=0}^{L-1}
  \intc_i(T) = \bar{\intc}(L,T).
\end{equation}
In the limit of large $T \gg L^2$, the
self-diffusion coefficient $\mathcal{D}_s(\rhobar)$ can be defined
through the mean-square tagged particle displacement of the
$\alpha^{th}$ particle as given below,
\begin{equation}
  \label{eq:self diff coeff}
  \angular{X^2_\alpha(T)} \simeq 2 \mathcal{D}_s(\rhobar) T.
\end{equation}
To compute the lhs of the above equation, we write the variance of the
sum $\sum_{\alpha=1}^N X_\alpha(T)$ as
\begin{align}
  \label{eq:sum of all displacement variance}
  \angular{\qty[\sum\limits_{\alpha=1}^N X_\alpha(T)]^2}
  =&
  \sum\limits_{\alpha, n} \sum\limits_{\alpha^\prime, {n^\prime}}
  \angular{\delta X_\alpha(t_n) \delta
  X_{\alpha^\prime}(t_{n^\prime})} \nonumber \\
  =& 2a(\rhobar) LT,
\end{align}
where   $\delta X_\alpha(t)$ is the microscopic displacement of the
$\alpha^{th}$  particle in a  small time interval $(t_n,t_n+\delta t)$ 
\begin{equation}
  \label{eq:net displacement def}
  X_\alpha(T) = \sum\limits_{n} \delta X_\alpha(t_n),
\end{equation}
and we have used eq.\eqref{eq:full system intc variance} in the last
line of eq.\eqref{eq:sum of all displacement variance}.
Now using $\angular{\delta X_\alpha(t) \delta X_{\alpha^\prime}(t^\prime)} \simeq 0$
for  $t \neq t^\prime$ and therefore
$\angular{X_\alpha^2(T)} = \sum\limits_n \angular{\delta
  X_\alpha^2(t_n)}$, we get
\begin{align}
  \label{eq:tag par result}
  \angular{\qty[\sum\limits_{\alpha=1}^N X_\alpha(T)]^2}
  \simeq \sum\limits_{\alpha=1}^N \angular{X_\alpha^2(T)} =
  N \angular{X^2_\alpha(T)}.
\end{align}
Comparing eqs.\eqref{eq:sum of all displacement
  variance},~\eqref{eq:self diff coeff}  
and~\eqref{eq:tag par result}, we obtain the following relations,
\begin{equation}
  \label{self diffusion coeff form}
  \mathcal{D}_s(\rhobar) = \frac{a(\rhobar)}{\rhobar}
  = \frac{1}{\rhobar} \qty[\lim\limits_{L,T \rightarrow \infty}
  \frac{\angular{\bar{\intc}^2(L,T)}}{2 L T}],
\end{equation}
which connects the self-diffusion coefficient, activity and the 
space-time integrated current fluctuation. 
Alternatively, one can show the above relation using a slightly
different argument as follows. First we note that
$\angular{X_{\alpha}^2(T)} = \angular{N_{\alpha}^{(h)}(T)}$, where
$N_{\alpha}^{(h)}(T)$ is the total number of hops, performed by the
$\alpha^{th}$ particle up to time $T$~\cite{Cunha_2009}. Summing over
all partciles we obtain
\begin{align}
  \label{eq:tag particle displacement}
  \sum\limits_\alpha  \angular{X_{\alpha}^2(T)} =
  \sum\limits_\alpha \angular{N_{\alpha}^{(h)}(T)} = 2 \angular{N^{(tp)}(T)},
\end{align}
where $N^{(tp)}(T)$ is the total number of toppling in the whole
system up to time $T$ and we have used the fact that
$\sum\limits_\alpha N_{\alpha}^{(h)}(T) = 2 N^{(tp)}(T)$. Also in the limit
of large $T$, we have in the leading order of $T$,
\begin{align}
  \label{eq:toppling activity relation}
  \angular{N^{(tp)}(T)} \simeq a(\rhobar) T L,
\end{align}
where $a(\rhobar)$ is the activity at density $\rhobar$. By summing
eq.\eqref{eq:self diff coeff} over all particles,
\begin{align}
  \label{eq:tag particle prop}
  \sum_{\alpha} \angular{X_\alpha^2(T)} \simeq 2
\mathcal{D}_s(\rhobar) T N,
\end{align}
and, then by using eqs.~\eqref{eq:toppling activity relation} and \eqref{eq:tag particle
  prop} in eq.\eqref{eq:tag particle displacement},  we obtain 
the relation as given in eq.\eqref{self diffusion coeff form}.

\begin{figure}[H]
  \centering
  \subfloat[\label{fig:tagged particle realizations}]{\includegraphics[scale=0.75]{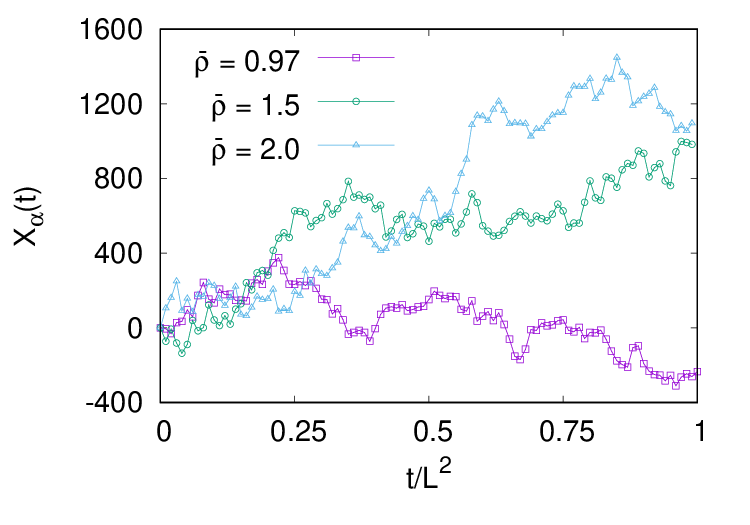}}\\
  \subfloat[\label{fig:diffusivities}]{\includegraphics[scale=0.75]{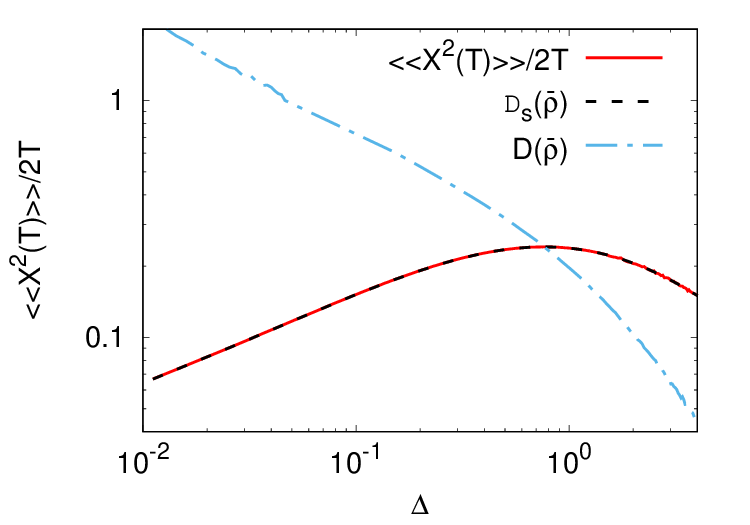}}
  \caption{
    In the top panel, we display the typical space-time trajectories for three tagged
    particles corresponding to the densities $\rho=0.97$ (violet
    square points), $1.5$ (green circular points), $2.0$ (blue triangular points)
    respectively with the scaled time axis.
    In the bottom panel, we plot the mean-square fluctuation of
    tagged particle displacement up to time $T$ (solid red line) as
    a function of relative density $\Delta$, where the double angular braces $\angular{\angular{X^2(T)}} = \sum_\alpha \angular{X_\alpha^2(T)}/N$
    denote the average over  trajectories as well as particles. Simulations (solid red line) show 
    excellent agreement  with the theoretically obtained self-diffusion coefficient 
   $\mathcal{D}_s(\rhobar)$ (dashed black line) as in 
    eq.\eqref{self diffusion coeff form}. In the same panel, we
    also plot the bulk-diffusion coefficient $D(\rhobar)=a'(\rhobar)$ as a function of
    $\Delta=\rhobar-\rho_c$ (dot-dashed blue line), using eq.\eqref{eq:D and a} from
    simulation, which has a contrasting behavior as compared to the self-diffusion coefficient
    $\mathcal{D}_s(\rhobar)$.} 
  \label{fig:tagged particle}
\end{figure}
In fig.~\ref{fig:tagged particle realizations},  we plot
typical trajectories of a particular tagged particle for
different densities $\rho=0.97$ (violet
    square points), $1.5$ (green circular points), $2.0$ (blue triangular points) as a function of scaled time.
In fig.~\ref{fig:diffusivities}, we plot the mean
square fluctuation of tagged particle displacement (solid red line) up
to time $T$, $\angular{\angular{X^2(T)}} / 2T$, as a function of
$\Delta = \rho-\rho_c$, where the double 
angular braces $\angular{\angular{X^2(T)}} = \sum_\alpha
\angular{X_\alpha^2(T)}/N$ denote the average over trajectories as
well as particles; in simulations, during the particle transfer at any
site $i$, two particles are chosen randomly 
from a particular stack. Our theoretical 
expression of the density dependent self-diffusion coefficient
$\mathcal{D}_s(\rho)$, as given in the equality
eq.\eqref{self diffusion coeff form}, is also plotted (dashed black
line); one can see an excellent agreement between the simulation
 and the theoretical prediction. 
For comparison, we also plot the bulk-diffusion coefficient
$D(\rhobar)$ (dot-dashed blue line), defined in eq.\eqref{eq:D and a}.
Here one should note that the self-diffusion coefficient 
$\mathcal{D}_s(\rhobar)$ and the bulk-diffusion coefficient $D(\rhobar)$
are, in principle, two different quantities and strikingly they have
quite contrasting  behaviors, especially near criticality.
Indeed, upon approaching criticality where activity decays as
$a(\rhobar) \sim (\rhobar - \rho_c)^\beta$ with $\beta < 1$,
the self-diffusion coefficient being the ratio of activity to
density (see eq.\eqref{self diffusion coeff form}) vanishes as
$\mathcal{D}_s(\rhobar) \sim (\rhobar - \rho_c)^\beta$ - exactly
in the same manner as the activity behaves near criticality, but the
bulk-diffusion coefficient being derivative of activity wrt the
density [see eq.\eqref{eq:D and a}] diverges as $D(\rhobar) \sim 1 / (\rhobar
-\rho_c)^{1-\beta}$~\cite{Chatterjee_PRE2018,Tapader_PRE2021}. Moreover, far
from criticality and in the limit of large density $\rhobar \gg 1$,
though the self-diffusion coefficient and the bulk-diffusion
coefficient both vanish, however they do so in different manners. In
that case, as the activity is expected to behave as $a(\rhobar) \simeq
1 - \text{const.}/\rhobar$, the self-diffusivity decays as
$\mathcal{D}_s(\rhobar) \sim 1/\rhobar$, but the bulk-diffusivity
decays much faster, $D(\rhobar) \sim 1/\rhobar^2$. Lastly, in the
active phase, where $\rhobar > \rho_c$, while the bulk-diffusion
coefficient $D(\rhobar)$ is a monotonically decreasing function of
density $\rhobar$ shown in fig.~\ref{fig:tagged particle}, the
self-diffusion coefficient $\mathcal{D}_s(\rhobar)$ is however a
non-monotonic function of $\rhobar$.

Importantly, unlike in the  symmetric simple
exclusion process where both the time-integrated bond current
and the tagged particle displacement fluctuations grow
sub-diffusively as $T^{1/2}$~\cite{Sadhu_2016, Masi_2002}, in the
conserved Manna 
sandpile only the  current fluctuation
grows sub-diffusively, whereas the tagged particle 
displacement fluctuation always grows linearly with time. This is
perhaps not surprising, given the fact that, in the Manna sandpile,
there are no restrictions in the  particle crossings, which are
otherwise not allowed in the symmetric exclusion process.

\subsection{Mass fluctuation and power spectrum}
\label{sec:mass fluctuation}

In the previous sections, we studied various properties of current
fluctuations in detail. Similarly, in this section, starting from the
microscopic update rules combined with the previously introduced truncation scheme, 
we shall derive various dynamic properties of mass fluctuations. The basic quantity is the
two-point dynamic correlation function
$C^{mm}_r(t,t^\prime) = \angular{m_0(t)m_r(t^\prime)} - \angular{m_0(t)}
\angular{m_r(t^\prime)}$. By using the microscopic update rules, we  write the time
evolution equation for $C_r^{mm}(t,0) \equiv C_r^{mm}(t)$ as
\begin{align}
  \label{eq:mass corr evolution}
  \dv{t} C_r^{mm}(t) =
  \sum\limits_k \Delta_{0,k} \angular{\hata_k(t) m_r(0)}.
\end{align}
Using the earlier truncation approximation eq.\eqref{eq:current approximation}, we
write the above equation as
\begin{align}
  \label{eq:approx mass eval eqn}
  \dv{t} C_r^{mm}(t)
  \simeq a^\prime(\rhobar) \sum\limits_k \Delta_{r,k} C^{mm}_{k}(t).
\end{align}
The solution of eq.\eqref{eq:approx mass eval eqn} can be written, by
using the Fourier representation, as
\begin{align}
  \label{eq:diff time mass corr sol}
  \tilde{C}_q^{mm}(t) \simeq e^{-a^\prime(\rhobar) \lambda_q t}
  \tilde{C}_q^{mm}(0),
\end{align}
where $\tilde{C}_q^{mm}$ is the Fourier transform of $C_r^{mm}$. The
equal-time mass correlation can
be solved by using the approximation eq.\eqref{eq:current approximation}
in eq.\eqref{eq:mass activity evl eqn same time} and we can write the
time evolution of $C_r^{mm}(t,t)$ in the steady state as
\begin{align}
  \label{eq:mass-mass corr steady state evl}
  \dv{t} C_r^{mm}(t,t) \simeq
  2 a^\prime(\rhobar) \sum\limits_k
  \Delta_{0,k} \angular{m_{k}m_{r}} + B_{r} = 0.
\end{align}
Similar to what was done earlier to solve eq.\eqref{eq:mass activity evl eqn same time}, the
above equation can be solved exactly using a generating function,
\begin{equation}
  \label{eq:mm generating function form}
  G(z) = \frac{1}{a^\prime(\rhobar)}
  \qty(\frac{3 a(\rhobar)}{2} - \frac{a(\rhobar)}{4} z).
\end{equation}
According to the above generating function, we have the steady-state correlations
$C_0^{mm} = \angular{m_0^2}-\rhobar^2 = 3a/2 a^\prime$, 
$C_1^{mm} = \angular{m_0 m_1}-\rhobar^2 = -a/4 a^\prime$ and all other correlations being zero. 
Thus we immediately arrive at a relation between the scaled subsystem-mass 
fluctuation and the activity,
\begin{equation}
  \label{eq:einstein equation derivation}
  \sigma^2(\rho) \equiv \lim\limits_{l \rightarrow \infty} 
  \frac{\angular{(\Delta M_l)^2}}{l}
  =  \sum\limits_{r=-\infty}^{r=\infty} C^{mm}_r =
 \frac{a(\rhobar)}{a^\prime(\rhobar)},
\end{equation}
where $\Delta M_l = M_l - \angular{M_l}$. Now, by using eqs. \eqref{eq:D and a} and \eqref{eq:sptm intjc limit}, the above identity can be recast into a nonequilibrium version of the Green-Kubo-like relation \cite{Chatterjee_PRE2018}, 
\begin{equation}
  \label{eq:GK}
  \sigma^2(\rhobar) = \frac{\sigma^2_Q(\rhobar)}{2 D(\rhobar)},
\end{equation}
connecting the (scaled) subsystem-mass fluctuation
$\sigma^2(\rhobar)$, the (scaled) subsysten-current fluctuation
$\sigma^2_Q(\rhobar)$ and the bulk-diffusion coefficient $D(\rhobar)$
(a slightly different form of the above relation is usually referred
as the Einstein relation in the literature~\cite{MFT_RMP2015,
  Chatterjee_PRE2018}). Remarkably, the fluctuation relation in
eq.~\eqref{eq:einstein equation derivation} implies that the scaled
subsystem mass fluctuation $\sigma^2(\rhobar)$ varies linearly with
the relative density $\Delta$, i.e., $\sigma^2(\rhobar) \sim
\Delta^{1-\delta}$ with $\delta=0$  \cite{Chatterjee_PRE2018};
interestingly, such behavior was indeed previously observed in
simulations \cite{Levine_PNAS2017} in a variant of the conserved Manna
sandpile, which is believed to be in the same universality class as
that studied here.

Next we write the solution of eq.\eqref{eq:diff time mass corr sol}
using the generating function in eq.\eqref{eq:mm generating function  form} as 
\begin{align}
  \label{eq:fourier diff time mm corr}
  \tilde{C}_q^{mm}(t) \simeq e^{-a^\prime(\rhobar) \lambda_q t}
  \frac{a(\rhobar)}{a^\prime(\rhobar)} \qty(1 + \frac{\lambda_q}{4}).
\end{align}
Finally, using the inverse Fourier transformation, we get,
\begin{equation}
  \label{eq:final diff time mm corr}
  C_r^{mm}(t) \simeq \frac{1}{L} \sum\limits_{q}
  e^{-\vb*{i} q r} e^{-a^\prime(\rhobar) \lambda_q t}
  \frac{a(\rhobar)}{a^\prime(\rhobar)} \qty(1 + \frac{\lambda_q}{4}).
\end{equation}
We now consider subsystem mass $M_l(t) = \sum_{r=0}^{l-1}
m_r(t)$ for $l < L$ and calculate the equal-time correlation function for mass
$C^{M_l M_l}(t,0) \equiv C^{M_l M_l}(t)$ by using the following expression,
\begin{align}
  \label{eq:subsystem mass corr def}
  C^{M_l M_l}(t) = l C_0^{mm}(t) + \sum\limits_{r=1}^{l-1} (l-r)
  \qty(C_r^{mm}(t) + C_{-r}^{mm}(t)).
\end{align}
Then by substituting eq.\eqref{eq:final diff time mm corr} in
eq.\eqref{eq:subsystem mass corr def}, we get the equal-time correlation for subsystem mass,
\begin{align}
  \label{eq:subsysmass corr result}
  C^{M_l M_l}(t) \simeq
  \frac{1}{L} \sum\limits_{q}
  e^{-a^\prime(\rhobar) \lambda_q t}
  \frac{a(\rhobar)}{a^\prime(\rhobar)} \qty(1 + \frac{\lambda_q}{4})
  \frac{\lambda_{lq}}{\lambda_q}.
\end{align}
For $t=0$, the correlation function $C^{M_l M_l} (0)$ is nothing but the equal-time
subsystem mass fluctuation, which can be written in the
large system size $L \rightarrow \infty$ limit as given below
\begin{align}
  \label{eq:cmm00}
  C^{M_l M_l}(0) = \angular{(M_l - \angular{M_l})^2} =
  \frac{a(\rhobar)}{a^\prime(\rhobar)} l
  \qty[1 + \frac{1}{2l}].
\end{align}
Then, by taking the large subsystem size $l \rightarrow \infty$ 
limit, where $1 \ll l \ll L$, we recover the  Einstein relation, already derived in
eq.\eqref{eq:einstein equation derivation}. Moreover the asymptotic form of 
eq.\eqref{eq:subsysmass corr result} can be written as
\begin{align}
  \label{eq:mass temp corr asymptotic}
  C^{M_l M_l}(t) -  C^{M_l M_l}(0)  \simeq  
-  \frac{2a(\rhobar)}{\sqrt{\pi a^\prime(\rhobar)}} t^{\frac{1}{2}}
\end{align}
for large time $1 \ll t \ll L^2$; see
appendix sec.~\ref{sec:temp corr subsys mass} for details..
Using the Fourier transform of eq.\eqref{eq:subsysmass corr result},
we write the power spectrum of the subsystem mass fluctuation as
\begin{align}
  \label{eq:subsys mass ps}
  S_{M}(f) = \lim\limits_{T \rightarrow \infty}
  \int\limits_{-T}^{T} \dd{t} C^{M_l M_l}(t) e^{2\pi \vb*{i} ft},
\end{align}
which can be written, by using eq.\eqref{eq:subsysmass corr result}, as 
\begin{align}
  \label{eq:subsys mass ps result}
  S_{M}(f) = \frac{1}{L} \sum\limits_{q}
  \frac{a(\rhobar)}{a^\prime(\rhobar)} \qty(1+\frac{\lambda_q}{4})
  \frac{2\lambda_q a^\prime(\rhobar)}{\lambda_q^2a^\prime(\rhobar)^2+4\pi^2f^2}
  \frac{\lambda_{lq}}{\lambda_q}.
\end{align}
In fig.~\ref{fig:mass exact ps}, we plot the power spectrum of the
subsystem 
mass fluctuation, obtained form simulation for $L=1000$ and $l=500$ in
solid lines, for various densities $\rhobar = 2.0$ (red line),
$\rhobar = 1.5$ (blue line), $\rhobar = 1.2$ (green line), $\rhobar =
1.0$ (purple line), $\rhobar = 0.97$ (orange line),
where the arrow accross the solid lines denotes the
ascending order of the density $\rhobar$. We compare the analytical
expression eq.\eqref{eq:subsys mass ps result}  (dotted black line)
with simulation result  for $\rhobar=2.0$ (solid red line), which is
in excellent agreement with theory. The asymptotic expression of the
power spectrum in eq.\eqref{eq:subsys mass ps result} can be
obtained by simplifying the integral as given below 
\begin{align}
  \label{eq:mass ps integral form}
  S_{M}(f)
  =& 4 a(\rhobar) \int\limits_{1/L}^{1/2} \dd{x} 
     \frac{\lambda(lx) \qty(1+\lambda(x) / 4)}
     {\lambda^2(x){a^\prime}^2(\rhobar)+4\pi^2f^2} 
     \nonumber \\
  \simeq&
          \frac{a(\rhobar)}{2\sqrt{\pi^3 a^\prime(\rhobar)}}
          f^{-\frac{3}{2}}.
\end{align}
Here, in the first step, we have replaced the sum in the rhs of
eq.\eqref{eq:subsys mass ps result} as an integral 
$(1/L) \sum_q \rightarrow \int_0^{2\pi} \dd{q}$ in the limit $L
\rightarrow \infty$ and we used $q = 2\pi x$, $\lambda(x) =
4\pi^2 x^2$ and eq.\eqref{eq:dimensionless y};
see appendix sec.~\ref{sec:subsysmass ps} for details. 
The above asymptotic form of the power spectrum can be used to
calculate the behavior near criticality by using the dimensional
scaling argument as performed before in eq.\eqref{eq:alpha def} where we write
$a/\sqrt{a^\prime} \sim f^{\mu}$. In other words,
near criticality, the decay of the power spectrum $S_{M}(f)$ as a function of frequency
$f$ becomes slower and is given by
\begin{align}
  \label{eq:subsys mass ps largel near criticality}
  S_{M}(f) \sim f^{-\psi_{M}},
\end{align}
where $\psi_{M} = 3/2 - \mu$. In fig.~\ref{fig:mass exact ps}, we
plot $S_{M}(f)$ as a function of frequency for 
densities $\rhobar = 2.0$ (red), $1.5$ (orange), $1.2$ (blue), $1.0$
(green), $0.97$ (violet); we observe that, as one approaches criticality,
the decay of the power spectrum indeed becomes slightly slower, in accordance with our
theoretical prediction in eq.\eqref{eq:subsys mass ps largel near criticality}.
The slower decay of the power spectrum up on approaching criticality can be physically understood from the current power spectrum as follows. Due to the slower temporal growth of the time-integrated bond current fluctuation (see eq.\eqref{eq:alpha def}), the near-critical
subsystem mass correlation also  decays slower as a function of time, i. e., 
\begin{align}
  \label{eq:near criticality mass corr decay}
 C^{M_l M_l}(t) -  C^{M_l M_l}(0)  \sim - t^{1/2-\mu},
\end{align}
which is  due to the fact that the  time-integrated current grows slower with time and
consequently the subsystem tends to retain a particular amount of mass for a much longer period. Indeed, this phenomenon can be thought of as the hyperuniformity  of mass fluctuations in the temporal domain - a dynamic hyperuniformity of mass fluctuation, analogous to
that of current fluctuations as described previously in eq.\eqref{eq:alpha def}.

\begin{figure}[H]
  \centering
  \includegraphics[scale=0.75]{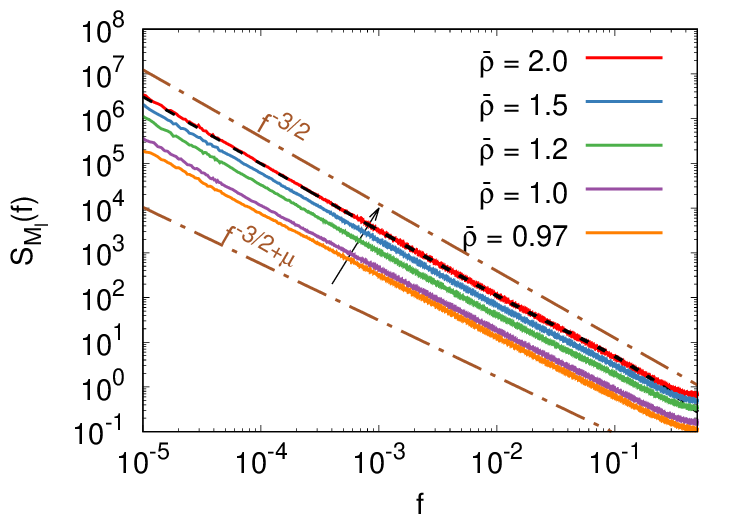}
  
  \caption{
    Power spectrum of subsystem mass fluctuations are plotted for
    $L=1000$ and $l=500$.
    The solid lines represent the simulation data for the densities
    $\rhobar = 2.0$ (red), $1.5$ (blue), $1.2$ (green), $1.0$ (purple),
    $0.97$ (orange) respectively. The top-most guiding line represents
    the $f^{-3/2}$ (away from criticality) behavior
    [eq. \eqref{eq:mass ps integral form}], whereas the bottom-most
    guiding line represents  $f^{-\Psi_{M}}$ (near criticality)
    behavior [eq.\eqref{eq:subsys mass ps largel near criticality}]
    where $\Psi_{M} = 3/2 - \mu \approx 1.26$. The dashed black line
    represents the theoretical result eq.\eqref{eq:subsys mass ps
      result} for $\rho = 2.0$. The arrow accross the
      solid lines signifies the ascending order of densities
      $\rhobar$.
  }
  \label{fig:mass exact ps}
\end{figure}

We note that the two exponents  $\psi_{\mathcal{J}}$ and $\psi_{M}$  for the current and mass power spectra defined in eqs.\eqref{eq:final largef ps curr near criticality} and \eqref{eq:subsys mass ps largel near criticality}, respectively, are  in fact related, due to the mass conservation as expressed in the continuity equation \eqref{eq:microscopic continuity  eq}. By using the Fourier transform of an observable $  A_r(t) = \int\limits_{-\infty}^{\infty} \dd{f} \sum\limits_q   e^{-2 \pi \vb*{i} f t} e^{-\vb*{i} q r}  \tilde{\mathcal{A}}_q(f)$, we can write eq.\eqref{eq:microscopic continuity eq} as $ -2\pi \vb*{i} f \tilde{\mathcal{M}}_q(f) =  \tilde{\mathcal{J}}_q(f)  \qty(e^{\vb*{i} q} - 1)$.
On a large scale $q \rightarrow 0$, we have
\begin{equation}
  \label{eq:mass and current power spectrum relation}
  S_{M}(f)  \sim f^{-2} S_{\mathcal{J}}(f),
\end{equation}
and therefore, from eqs.\eqref{eq:final largef ps curr near criticality}
and \eqref{eq:subsys mass ps largel near criticality}, we obtain the scaling relation
\begin{equation}
  \label{eq:mass and curr ps exponent relation}
  \psi_{\mathcal{J}} = 2 - \psi_{M}.
\end{equation}

\section{Summary and conclusions}
\label{sec:conclusion}

In this paper, we study the steady-state dynamical properties of current and mass in the active phase of the one-dimensional conserved Manna sandpile, and we establish a direct quantitative relationship between the system's static and dynamic properties. First starting with a microscopic dynamical description, we introduce a truncation scheme, that is approximate and is expected to be valid only for long (hydrodynamic) times, but allows us to theoretically investigate the time-dependent (two-point, unequal-time) correlation functions for current and mass, as well as the associated power spectra. 
In particular, we find that, in the thermodynamic limit, the two-point time-dependent correlation function for the (bond) current has a delta peak at time $t=0$ and, for time $t > 0$, the correlation is \textit{negative} and a long-ranged one, decaying as $t^{-(3/2 + \mu)}$. Far from criticality, we show that the exponent $\mu = 0$, resulting in a subdiffusive, $T^{1/2}$, growth of the variance of the cumulative (\textit{time-integrated}) current up to time $T$. This type of subdiffusive growth of temporal fluctuation, which has previously been obtained in symmetric simple exclusion processes \cite{Sadhu_2016}, is somewhat expected for diffusive systems with normal fluctuation properties, such as sandpiles away from criticality \cite{Tapader_PRE2021}. However the scenario changes near the critical point. Indeed, as one approaches criticality, the activity in the system vanishes, contributing to an anomalous suppression of the temporal current fluctuations and thus a positive value of the exponent $\mu > 0$, which has been expressed in terms of the standard static exponents [see eq.\eqref{eq:mu def}]; likewise, near criticality, the power spectrum of current at low frequency $f$ varies as $f^{1/2+\mu}$.
A similar argument can be made for the temporal subsystem-mass fluctuation, which is induced by the boundary currents and is also suppressed near criticality because the current fluctuation is suppressed. The anomalously reduced mass fluctuation is manifested in the corresponding power spectrum, which, at low frequency and near criticality, varies as $f^{-3/2+\mu}$ with  $\mu > 0$; on the other hand, far from criticality, the exponent $\mu=0$, implying $f^{-3/2}$ power spectrum, expected in a normal diffusive system.
We also derive, within our theory, a nonequilibrium version of the Green-Kubo-like fluctuation-response relation [see eq.\eqref{eq:GK}], or the Einstein relation \cite{MFT_RMP2015, Chatterjee_PRE2018}, which connects dynamic and static fluctuations in the system.
Indeed our theoretical analysis suggests that, with appropriate
(diffusive) rescaling of space and time, the fluctuation properties of
the Manna sandpile should be governed by a continuum fluctuating
hydrodynamic description as formulated in the recently developed
macroscopic fluctuation theory for diffusive
systems~\cite{MFT_RMP2015, Chatterjee_PRE2018}. 

We finally investigate the mean-square displacement of tagged-particles and show that the self-diffusion coefficient for an individual tagged particle is identically equal to the ratio of the activity to density [see first equality in eq.\eqref{self diffusion coeff form}]. The identity readily explains a previous simulation observation of Ref.~\cite{Cunha_2009} that the self-diffusion coefficient in the Manna sandpile vanishes in precisely the same fashion as the activity does upon approaching criticality. Notably, the near-critical behavior of the self-diffusion coefficient differs markedly from that of the bulk-diffusion coefficient, which was previously identified in Refs.~\cite{Chatterjee_PRE2018, Tapader_PRE2021} as the derivative $a^\prime(\rhobar)$ of the activity wrt density $\rhobar$ and clearly diverges near criticality.
Furthermore, while the bulk-diffusion coefficient is a monotonic function of density, the self-diffusion coefficient is a non-monotonic one.
Interestingly, the self-diffusion coefficient can be related to the current fluctuation in the system as expressed in the second equality in eq.\eqref{self diffusion coeff form}.

Our findings are reminiscent of somewhat similar observations of \textit{dynamic hyperuniformity}, where the existence of anti-correlations in the temporal fluctuations in sandpiles were pointed out ~\cite{Hwa_1989, Garcia-Millan_2018}. However, until recently \cite{Chatterjee_PRE2018}, the precise relationship between dynamic and static fluctuations, such as that between mass and current fluctuations, was unknown, and is encoded in the Green-Kubo-like fluctuation relation as derived here in eq.\eqref{eq:GK}. The relationship demonstrates that there are indeed two mechanisms responsible for the vanishing of mass fluctuation near criticality: Physically, the anomalously suppressed current fluctuation, combined with the diverging bulk-diffusion coefficient, near criticality result in the vanishing, or hyperuniform, density fluctuation observed recently in the conserved Manna sandpiles ~\cite{Levine_PRL1_2015}.

Our results are in fact a consequence of the mass conservation, as reflected in the scaling relation \eqref{eq:mass and curr ps exponent relation}, and are thus expected to be applicable in a broad class of conserved sandpiles. As our analysis suggests, the anomalous suppression of current fluctuations near criticality could be a generic feature of \textit{hyperuniform} state of matter and should serve as the {\it dynamical} signature of such states, which have been observed in similar other systems in the recent past \cite{Chaikin_2008, Levine_PNAS2017}. 
In particular, our findings could help in determining the precise dynamical nature of the off- and near-critical states in sandpiles by shedding light on the microscopic dynamical origin of long-ranged temporal correlations in these systems.

\begin{acknowledgments}
  We thank Deepak Dhar for useful discussions and comments on the manuscript.
  A.M. acknowledges financial support from the the Department of
  Science and Technology, India [Fellowship No. DST/INSPIRE
  Fellowship/2017/IF170275] for part of the work carried out under his
  senior research fellowship.  We acknowledge the
  Thematic Unit of Excellence on Computational Materials Science, funded
  by the Department of Science and Technology, India for computational
  facility.
\end{acknowledgments}

\bibliography{new_reference}

\begin{thebibliography}{62}
\bibitem[{ani()}]{anirbanthesis}
\bibinfo{journal}{The details of the results will be presented elsewhere}
  (????).
\end{thebibliography}
\newpage

\appendix

\begin{widetext}

\section{Some algebraic identities and special integrals}
\label{sec:identities and integrals}

We can deduce several algebric properties of $\lambda_n$, which are
the following,
\begin{align}
  \label{eq:lambda prop0}
  \sum\limits_{n=1}^{L-1} \frac{\lambda_{nl}}{\lambda_n^2} =
  \frac{1}{12} l (l-L) \left(l^2-l L-2\right);
\end{align}

\begin{align}
  \label{eq:lambda prop1}
  \sum\limits_{n=1}^{L-1} \frac{1}{\lambda_n} = \frac{L^2-1}{12};
\end{align}

\begin{align}
  \label{eq:lambda prop2}
  \sum\limits_{n=1}^{L-1} \frac{\lambda_{nr}}{\lambda_n} = r(L-r);
\end{align}

\begin{align}
  \label{eq:lambda prop3}
  \sum\limits_{r=1}^{l-1} 2(l-r)(2-\lambda_{rn})
  = 2\qty(\frac{\lambda_{ln}-l\lambda_n}{\lambda_n}).
\end{align}

\begin{align}
  \label{eq:lambda prop4}
  \sum\limits_{n=1}^{L-1} \lambda_{nl} = 2L, \hspace{4pt} \text{for}
  \hspace{4pt} l=1,2,\ldots.
\end{align}
Eq.\eqref{eq:lambda prop1} is a special case of
eq.\eqref{eq:lambda prop0} for $l=1$.

The integrals appeared in the context of asymptotic analysis, i.e., in
eq.\eqref{eq:ps curr asymptotic} and later in
eq.\eqref{eq:subsysmass ps large l leading} have very generic
solutions in terms of hypergeometric functions. Generically, we can
write those integrals in the following form,
\begin{align}
  \label{eq:special integral}
  I(y) = \int\limits_0^y \dd{z} \frac{z^{-k}}{1+z} =
  \frac{y^{1-k} \, _2F_1(1,1-k;2-k;-y)}{1-k},
\end{align}
where $_2F_1$ is the hypergeometric
function~\cite{gasper2004}, defined as
\begin{align}
  \label{eq:hypergeometric function}
  _2F_1(a,b;c;z) = 
  \frac{\Gamma(c)}{\Gamma(b) \Gamma(c-b)}
  \int\limits_0^1 \dd{t} t^{b-1} (1-t)^{c-b-1} (1-tz)^{-a}.
\end{align}
In the limit of ${y \rightarrow \infty}$, we have,
\begin{align}
  \label{eq:special integral limit1}
  \lim\limits_{y \rightarrow \infty} I(y) = \sqrt{2} \pi
  \hspace{5pt} \text{for} \hspace{5pt} k = \frac{1}{4}, \frac{3}{4}.
\end{align}
We have used the above result in
eqs.~\eqref{eq:ps curr asymptotic},~\eqref{eq:subsysmass ps large l
  leading}  to obtain the asymptotic for of the
power spectrum of current and subsystem mass respectively.

\section{Asymptotic analysis}
\label{sec:asymptotic analysis}

In this section, we provide the calculation details of the results
eq.\eqref{eq:integrated current asymptotic},~\eqref{eq:integrated
current correlation},~\eqref{eq:inst curr asymptotic},~\eqref{eq:sptm
large time fluc},~\eqref{eq:mass temp corr asymptotic}, presented in
the main text.

\subsection{Time-integrated bond current fluctuation}
\label{sec:intj fluc asymptotic}

We now derive the asymptotic approximation of the time-integrated bond
current correlation using 
eq.\eqref{eq:integrated current asymptotic}. We can write the
unequal-time time-integrated bond current correlation as
\begin{align}
  \label{eq:unequal time bond current correlation}
  C^{\intc \intc}_0(t,t^\prime)
  =& \frac{2a(\rhobar)}{L} t^\prime +
  2 a(\rhobar) a^\prime(\rhobar) \frac{1}{L} \sum\limits_q
  \frac{1 - e^{-\lambda_q a^\prime(\rhobar) t^\prime}}{\lambda_q^2
     {a^\prime}^2} \lambda_q \qty(1 + \frac{\lambda_q}{4}) -
     \nonumber \\
   & a(\rhobar) a^\prime(\rhobar) \frac{1}{L} \sum\limits_q
  \frac{1 - e^{-\lambda_q a^\prime(\rhobar) t^\prime} +
  e^{-\lambda_q a^\prime(\rhobar) t}
  - e^{-\lambda_q a^\prime(\rhobar) (t - t^\prime)}}{\lambda_q^2
  {a^\prime}^2} \lambda_q \qty(1 + \frac{\lambda_q}{4})
\end{align}
In the infinite system size limit $L \rightarrow \infty$, we write the
above sum in the following integral form,
\begin{align}
  \label{eq:intj corr simplified}
     C^{\intc\intc}_0(t,t^\prime) \simeq
  & 4a^\prime(\rhobar)a(\rhobar)  \int\limits_{0}^{1/2} \dd{x}
    \frac{1 - e^{-\lambda(x)a^\prime(\rhobar) t^\prime}}{\lambda(x)^2
    a^\prime(\rhobar)^2} 
    \lambda(x) \qty(1+\frac{\lambda(x)}{4}) 
    \nonumber \\
    & -2a^\prime(\rhobar)a(\rhobar)  \int\limits_{0}^{1/2} \dd{x}
      \frac{1 - e^{-\lambda(x) a^\prime(\rhobar) t^\prime} +
      e^{-\lambda(x) a^\prime(\rhobar) t} - e^{-\lambda(x) 
      a^\prime(\rhobar) (t-t^\prime)}} {\lambda^2(x) {a^\prime}^2(\rhobar)}
      \lambda(x) \qty(1+\frac{\lambda(x)}{4}).
\end{align}
Note that the integral in the above equation can be expressed in terms of an
integral of the form as given below,
\begin{align}
  \label{eq:intj corr approx1}
  a(\rhobar) a^\prime(\rhobar)\int\limits_{0}^{1/2} \dd{x}
     \frac{1-e^{-\lambda(x)a^\prime(\rhobar) t^\prime}}{\lambda^2(x)
  {a^\prime}^2(\rhobar)} 
  \lambda(x) \qty(1+\frac{\lambda(x)}{4}) \simeq
  \frac{a(\rhobar)\sqrt{t^\prime}}{2\sqrt{\pi a^\prime(\rhobar)}},
\end{align}
where, for $t \gg 1$, we have defined 
$x = \qty({y}/{4\pi^2 a^\prime(\rhobar) t^\prime})^{1/2}$ and used
$\lambda(x) \simeq 4\pi^2 x^2$ and $\int\limits_0^\infty \dd{y}
y^{-3/2}(1-e^{-y}) = 2\sqrt{\pi}$, 
to explicitly calculate the integral.
Using eq.\eqref{eq:intj corr approx1} in each of the relevant terms of
the rhs in eq.\eqref{eq:intj corr simplified} and then after some
straightforward algebraic manipulations, we obtain the following
asymptotic form of the time-dependent integrated bond-current correlation,
\begin{align}
  \label{eq:intj shorttime asymptotic}
  C^{\intc\intc}_0(t,t^\prime) \simeq
  \frac{a(\rhobar)}{\sqrt{\pi a^\prime(\rhobar)}}\qty(\sqrt{t} +
  \sqrt{t^\prime} - \sqrt{\abs{t-t^\prime}}).
\end{align}
Now, by putting $t^\prime = t \equiv T$, the above asymptotic leads to the
first part (i.e., corresponding to the limit $1 \ll T \ll L^2$) of
eq.\eqref{eq:integrated current asymptotic} in the main text.

\subsection{Time-dependent instantaneous current correlation}
\label{sec:inst curr fluc asymptotics}

The steady state unequal-time correlation of instantaneous bond current
$C^{\mathcal{J} \mathcal{J}}_0(t,0)  =
\angular{\mathcal{J}_0(0) \mathcal{J}_r(t)} -
\angular{\mathcal{J}_0(0)} \angular{\mathcal{J}_r(0)} $, for 
$t\geq 0$, is given by the following expression as derived in main
text (see eq.\eqref{eq:t>s curr corr})
\begin{align}
    \label{eq:curr self cumulant}
  C^{\mathcal{J} \mathcal{J}}_0(t,0) = \delta(t)3a(\rhobar)
  - a^\prime(\rhobar)a(\rhobar)\frac{1}{L}
  \sum\limits_{q}
  e^{-a^\prime(\rhobar)\lambda_q t} 
  \lambda_q \qty(1+\frac{\lambda_q}{4}).
\end{align}
First, we perform the time integral in a finite time domain
$[-T,T]$ as given below,
\begin{align}
  \label{eq:integrated current correlation appendix}
  \int\limits_{-T}^{T} C^{\mathcal{J} \mathcal{J}}_0(t,0) \dd{t} =
  \Gamma_0(\rhobar) - \frac{2a(\rhobar)}{L} \sum\limits_q
  \qty(1+\frac{\lambda_q}{4}) + 2a(\rhobar)
  \qty[\frac{1}{L} \sum\limits_q e^{-\lambda_q a^\prime(\rhobar) T}
  \qty(1+\frac{\lambda_q}{4}) ],
\end{align}
which, 
using the relations eq.\eqref{eq:lambda prop4} and
$\Gamma_0(\rhobar) = 3a(\rhobar)$ as in the main text in
eq.\eqref{eq:gamma properties}, we simplify the above sum as
\begin{align}
  \label{eq:integrated current correlation appendix2}
  \int\limits_{-T}^{T} C^{\mathcal{J} \mathcal{J}}_0(t,0) \dd{t} =
  \frac{2a(\rhobar)}{L}  + 2a(\rhobar)
  \qty[\frac{1}{L} \sum\limits_q e^{-\lambda_q a^\prime(\rhobar) T}
  \qty(1+\frac{\lambda_q}{4}) ].
\end{align} 
Now, first taking the infinite-system size limit, i.e., the limit $L
\rightarrow \infty$, we can further write the sum as an 
integral,
\begin{align}
  \label{eq:integrated current correlation appendix3}
  \lim\limits_{L \rightarrow \infty}\int\limits_{-T}^{T}
  C^{\mathcal{J} \mathcal{J}}_0(t,0) \dd{t} = 
  4a(\rhobar) \int\limits_0^{1/2} \dd{x}
  e^{-\lambda(x) a^\prime(\rhobar) T}
  \qty(1+\frac{\lambda(x)}{4}),
\end{align} 
where $  q = 2\pi x$.
Now using $\lambda(x) \simeq 4\pi^2 x^2$, $\int_0^\infty \dd{y} e^{-y}
y^{-1/2} = \sqrt{\pi}$ and a variable
transformation $x = y^{1/2} / 2\pi \sqrt{a^\prime(\rhobar) T}$,
we can explicitly calculate the integral as in
eq.\eqref{eq:integrated current correlation appendix3} as
\begin{align}
  \label{eq:integrated current correlation appendix4}
  \lim\limits_{L \rightarrow \infty}\int\limits_{-T}^{T}
  C^{\mathcal{J} \mathcal{J}}_0(t,0) \dd{t} \simeq 
  \frac{a(\rhobar)}{\pi \sqrt{a^\prime(\rhobar) T}}
  \int\limits_0^{\infty} \dd{y} e^{-y} y^{-1/2} =
  \frac{a(\rhobar)}{\sqrt{\pi a^\prime(\rhobar) }} T^{-1/2},
\end{align}
which is the result in the main text in
eq.\eqref{eq:integrated current correlation}.
Finally, by taking the limit $T \rightarrow \infty$, we get
\begin{align}
  \label{eq:integrated current correlation appendix4}
  \int\limits_{-\infty}^{\infty}
  C^{\mathcal{J} \mathcal{J}}_0(t,0) \dd{t} = 0,
\end{align}
which is the result presented in main text in
eq.\eqref{eq:infinte sptm vol cur corr}.

Similarly, we can find the asymptotic form of
$c^{\mathcal{J} \mathcal{J}}_0(t,0)$ presented in main text in
eq.\eqref{eq:inst curr asymptotic}.
In the limit
$L \rightarrow \infty$, the temporal current correlation 
$c^{\mathcal{J} \mathcal{J}}_0(t,0)$ for $t > 0$ can be written as an
integral, as given below
\begin{align}
  \label{eq:continuous limit curr self cumulant}
  c^{\mathcal{J} \mathcal{J}}_0(t,0) \simeq 
  - 2a^\prime(\rhobar)a(\rhobar) 
  \int\limits_{0}^{1/2} \dd{x}
  e^{-a^\prime(\rhobar)\lambda(x)t} 
  \lambda(x) \qty(1+\frac{\lambda(x)}{4}).
\end{align}
Now, again using $\lambda(x) \simeq 4\pi^2 x^2$ and a variable
transformation $x = y^{1/2} / 2\pi \sqrt{a^\prime(\rhobar) T}$,
we calculate the above integral as
\begin{align}
  \label{eq:initial integral form}
  c^{\mathcal{J} \mathcal{J}}_0(t,0)
  \simeq - \frac{a(\rhobar)}{\sqrt{4\pi^2 a^\prime(\rhobar)}} t^{-3/2} 
  \int\limits_0^{\infty}
     \dd{y} \sqrt{y} e^{-y} ,
\end{align}
where we have ignored the subleading term $\order{t^{-5/2}}$.
Finally, using  $\int\limits_0^{\infty} 
\dd{y} \sqrt{y} e^{-y} = \sqrt{\pi}/2$, we get the
result presented in main text in eq.\eqref{eq:inst curr asymptotic},
\begin{align}
  \label{eq:final self cumulant}
    c^{\mathcal{J} \mathcal{J}}_0(t,0) \simeq
  - \frac{a(\rhobar)}{4\sqrt{\pi a^\prime(\rhobar)}} t^{-3/2}.
\end{align}
 
  \subsection{Spacetime-integrted current fluctuation}
  \label{sec:sptm integrated current}
  
  Here we derive the asymptotic dependence of 
  eq.\eqref{eq:final sptm QiQj fluctuations} on subsystem size $l$ and
  time $T$ (see eq.\eqref{eq:sptm large time fluc}); first by taking
  the limit $T \gg 1$ and $l \gg 1$ and then 
  followd by the reverse order of limit $l \gg 1$ and $T \gg 1$. In
  both cases, we take infinite system size limit $L \rightarrow
  \infty$, therefore $l / L \rightarrow 0$ is
  always satisfy.

  \subsubsection{Case I: $T \gg 1$, $l \gg 1$}
  \label{sec:sptm intj asymptotic I}

  In this case, we write
  eq.\eqref{eq:final sptm QiQj fluctuations} in the following
  simplified form,
  \begin{align}
    \label{eq:simplified sptm intj fluc}
    \angular{\bar{Q}^2(l,T)} = \frac{2 a(\rhobar) T
    l^2}{L}
    + 2a(\rhobar) a^\prime(\rhobar) \frac{1}{L}
    \sum\limits_q \frac{1 - e^{-a^\prime(\rhobar) \lambda_q
    T}}{\lambda^2_q {a^\prime}^2} \qty(1 + \frac{\lambda_q}{4})
    \lambda_{ql}.
  \end{align}
  In the limit $L \rightarrow \infty$, the sum in the above equation can be
  converted in to the following integral,
  \begin{align}
    \label{eq:sptm intj fluc integral1}
    \angular{\bar{Q}^2(l,T)} \simeq
    4 a(\rhobar) a^\prime(\rhobar) \int\limits_0^{1/2} \dd{x}
    \frac{1 - e^{-a^\prime(\rhobar) \lambda(x) T}}{\lambda^2(x)
    {a^\prime}^2} \qty(1 + \frac{\lambda(x)}{4}) \lambda(lx).
  \end{align}
  Using the approximation $\lambda(lx) \simeq 4\pi^2 l^2x^2$ for
  finite subsystem size $l$ and a
  variable transformation $x = y^{1/2} / 2\pi \sqrt{a^\prime(\rhobar)
    T}$, we get
  \begin{align}
    \label{eq:sptm intj fluc integral2}
    \frac{1}{lT} \angular{\bar{Q}^2(l,T)} \simeq
    \frac{a(\rhobar)}{\pi \sqrt{a^\prime(\rhobar)}} \frac{l}{\sqrt{T}} 
    \int\limits_0^\infty \dd{y} (1 - e^{-y}) y^{-3/2}
    = \frac{2 a(\rhobar)}{\sqrt{\pi a^\prime(\rhobar)}}
    \frac{l}{\sqrt{T}},
  \end{align}
  where we use   $\int_0^\infty \dd{y} (1-e^{-y}) y^{-3/2} =
  2\sqrt{\pi}$. This result appeared in eq.\eqref{eq:sptm large time
    fluc} of the main-text.

  \subsubsection{Case II: $l \gg 1$, $T \gg 1$}
  \label{sec:sptm intj asymptotic II}

  To compute the asymptotic form in this limit,
  we use the approximation $\lambda(lx) \simeq 2$ to write
  eq.\eqref{eq:final sptm QiQj fluctuations} in the following form,
  \begin{align}
    \label{eq:approx sptm QiQj fluctuations}
   \frac{1}{lT} \angular{\bar{Q}^2(l,T)}_c
    \simeq
    2a(\rhobar) + \frac{a(\rhobar)}{l} -
    \frac{8a(\rhobar)a^\prime(\rhobar)}{lT} \int\limits_{0}^{1/2}  
    \frac{a^\prime(\rhobar)\lambda(x) T -1+\exp\qty(-\lambda(x)
    a^\prime(\rhobar)T)}{\lambda^2(x) {a^\prime}^2(\rhobar)} 
    \qty(1+\frac{\lambda(x)}{4}).
  \end{align}
  Again using the variable transform $x = y^{1/2} / 2\pi
  \sqrt{a^\prime(\rhobar) T}$ and
  $\int_0^\infty \dd{y} \qty(y - 1 + e^{-y}) y^{-5/2} = 4\sqrt{\pi} /
  3$, we get   
  \begin{align}
    \label{eq:sptm intj asymptotic}
    \frac{1}{lT} \angular{\bar{Q}^2(l,T)}_c
    \simeq&
            2a(\rhobar) + \frac{a(\rhobar)}{l} -
            \frac{2a(\rhobar)a^\prime(\rhobar)}{lT}
            \int\limits_{0}^{\infty} \dd{y} \qty(y -1+e^{-y})
            y^{-5/2} \frac{T^{3/2}}{\pi \sqrt{a^\prime(\rhobar)}} \nonumber \\
    =& 2a(\rhobar) + \frac{a(\rhobar)}{l} -
       \frac{8 a(\rhobar)}{3}  \sqrt{\frac{a^\prime(\rhobar)}{\pi}}
       \frac{\sqrt{T}}{l},
  \end{align}
  which also appeared in eq.\eqref{eq:sptm large time fluc}.

  \subsection{Temporal-correlation of subsystem mass}
  \label{sec:temp corr subsys mass}
  
  The asymptotic form of subsystem mass temporal correlation $C^{M_l
    M_l}(t,0)$, appeared in eq.\eqref{eq:mass temp corr asymptotic} is
  derived here. At $t=0$, $C^{M_l M_l}(t,0)$ is maximum and after that,
  it decays as a function of time $t$. So, in order to extract the
  temporal dependence of $C^{M_l M_l}(t,0)$, we write 
  eq.\eqref{eq:subsysmass corr result} as
  \begin{align}
    \label{eq:subsystem mass corr appendix}
    C^{M_l M_l}(0,0) - C^{M_l M_l}(t,0) =
    \frac{1}{L} \sum\limits_q
    \qty(1 - e^{-\lambda_q a^\prime(\rhobar) t}) \qty(1 +
    \frac{\lambda_q}{4}) \frac{\lambda_{lq}}{\lambda_q},
  \end{align}
  which, in the limit $L \rightarrow \infty$, $l \gg 1$, can be
  written as 
  \begin{align}
    \label{eq:mass corr integral}
    C^{M_l M_l}(0,0) - C^{M_l M_l}(t,0) \simeq 4\int\limits_{0}^{1/2} \dd{x}
    \frac{a(\rhobar)}{a^\prime(\rhobar)}
    \qty(1 - e^{-\lambda(x) a^\prime t})
    \qty(1 + \frac{\lambda(x)}{4}) 
    \frac{1}{\lambda(x)}.
  \end{align}
  The above equation can be further simplified using the approximation
  and exact results of the previous
  sec.~\ref{sec:sptm intj asymptotic I} and we get
  \begin{align}
    \label{eq:mass corr final asymptotic}
    C^{M_l M_l}(0,0) - C^{M_l M_l}(t,0) =
    \frac{a(\rhobar)}{\sqrt{a^\prime(\rhobar) \pi^2}} t^{\frac{1}{2}}
    \int_0^\infty \dd{y} y^{-\frac{3}{2}} \qty(1-e^{-y})
    = \frac{2a(\rhobar)}{\sqrt{\pi a^{\prime}(\rhobar)}} t^{\frac{1}{2}}.
  \end{align}
  Thus we derived  eq.\eqref{eq:mass temp corr asymptotic} of the
  main-text.

\subsection{Power spectrum of subsystem mass fluctuation}
\label{sec:subsysmass ps}

In order to find the asymptotic form of the power spectrum of
subsystem mass fluctuation $S_{M_l}(f)$, in the limit $L \rightarrow
\infty$ and $l \gg 1$, we write 
eq.\eqref{eq:subsys mass ps result} as
\begin{align}
  \label{eq:continuous subsys mass ps}
  S_{M_l}(f)
  \simeq \frac{8 a(\rhobar)}{a^\prime(\rhobar)} \int\limits_{0}^{1/2}
  \frac{\lambda(x) a^\prime(\rhobar)}
  {\lambda^2(x) {a^\prime}^2(\rhobar) + 4\pi^2 f^2}
  \qty(1 + \frac{\lambda(x)}{4}) \frac{1}{\lambda(x)}.
\end{align}
Using the variable transform defined in
eq.\eqref{eq:dimensionless y} and $\lambda(x) = 4 \pi^2 x^2$, the
above equation can be written in the following  simplified form,
\begin{align}
  \label{eq:subsysmass ps large l leading}
  \tilde{S}_{M_l}(f)
  & \simeq \frac{a}{2\pi^2 \sqrt{2\pi a^\prime(\rhobar)}} f^{-3/2} \times
    \int\limits_0^\infty \frac{y^{-3/4}}{1+y} \dd{y} =
    \frac{a}{2 \sqrt{\pi^3 a^\prime(\rhobar)}} f^{-3/2},
\end{align}
where, we used eq.\eqref{eq:special integral limit1} and ignored the
term of $\order{f^{1/2}}$. Thus we derive 
eq.\eqref{eq:subsys mass ps result}, appeared in the main-text.

\section{Evolution equations of correlation functions}
\label{sec:evl eqns}

\subsection{Different time current-current correlation}
\label{sec:diff time intc intc corr}

The stochastic update rules for the different-time and different-space
product function of currents
$\intc_i(t) \intc_j(t^\prime)$,  can be written for $t > t^\prime$ as
\begin{align}
  \label{eq:diff intc intc update rules}
  \intc_i(t+\dd{t}) \intc_j(t^\prime) =
  \begin{cases}
    \mathbf{events}
    & \mathbf{probabilities} \\
    \qty(\intc_i(t) + 1) \intc_j(t^\prime)
    & \frac{1}{2} \hata_i(t) \dd{t} \\
    \qty(\intc_i(t) + 2) \intc_j(t^\prime)
    & \frac{1}{4} \hata_i(t) \dd{t} \\
    \qty(\intc_i(t) - 1) \intc_j(t^\prime)
    & \frac{1}{2} \hata_{i+1}(t) \dd{t} \\
    \qty(\intc_i(t) - 2) \intc_j(t^\prime)
    & \frac{1}{4} \hata_{i+1}(t) \dd{t} \\
    \intc_i(t) \intc_j(t^\prime)
    & 1 - \Sigma \dd{t},
  \end{cases}
\end{align}
where $\Sigma = 3\qty(\hata_i(t) + \hata_{i+1}(t))/4$.
Using these rules, we write the evolution equation of the two
point current-current correlation function as
\begin{align}
  \label{eq:diff time intc intc evl eqn}
  \pdv{t} C^{\intc \intc}_r (t,t^\prime) =
  \angular{\qty[\hata_0(t) -\hata_1(t)] \intc_r(t^\prime)} =
  \qty[C_r^{\hata \intc}(t,t^\prime) - C_{r-1}^{\hata \intc}(t,t^\prime)],
\end{align}
which has appeared in
eq.\eqref{eq:diff time intc intc corr update eqn} in the main-text.

\subsection{Different time mass-current correlation}
\label{sec:diff time mass intc corr}

The stochastic update rules for the different-time and different-space
product function of mass and current
$m_i(t) \intc_j(t^\prime)$,  can be written for $t > t^\prime$ as
\begin{align}
  \label{eq:diff time mass intc update rules}
  m_i(t+\dd{t}) \intc_j(t^\prime) =
  \begin{cases}
    \qty(m_i(t) +1) \intc_j(t^\prime) & \frac{1}{2}\hata_{i+1}(t) dt \\
    \qty(m_i(t) +1) \intc_j(t^\prime) & \frac{1}{2}\hata_{i-1}(t) dt \\
    \qty(m_i(t) +1) \intc_j(t^\prime) & \frac{1}{4}\hata_{i+1}(t) dt \\
    \qty(m_i(t) +2) \intc_j(t^\prime) & \frac{1}{4}\hata_{i-1}(t) dt \\
    \qty(m_i(t) -2) \intc_j(t^\prime) & \hata_{i}(t) dt \\
    m_i(t) \intc_j(t^\prime) & [1-\Sigma dt],
  \end{cases}
\end{align}
where $\Sigma = 3 \qty(\hata_i(t) + \hata_{i+1}(t))/4 + \hata_i(t)$.
Using these rules, we write the evolution equation of the two
point mass-current correlation function as
\begin{align}
  \label{eq:diff time intc intc evl eqn}
  \pdv{t} C^{m \intc}_r (t,t^\prime) =
  \angular{\qty[\hata_{L-1}(t) - 2\hata_0(t) +\hata_1(t)]
  \intc_r(t^\prime)}
  \simeq a^\prime(\rhobar) \sum\limits_k
  \Delta_{r,k}C^{m\intc}_k(t,t^\prime),
\end{align}
which has appeared in the
eq.\eqref{eq:mass current evl eqn}.

\subsection{Equal time current-current correlation}
\label{sec:same time intc intc}

In the Manna sandpile, during each toppling two particles can hop to
each neighbouring site intependently and it may simulataneously create
current at two neighbouring bonds. In the following, we write the
update equation of the two point product function of  integrated current,
  \begin{equation}
  \label{eq:equal time intc intc}
  \intc_i(t+\dd{t}) \intc_j(t+\dd{t}) = \\
  \begin{cases}
    & \mathit{probabilities} \\
    \qty(\intc_i(t) - 1) \qty(\intc_j(t) + 1)
    & \frac{1}{2} \hata_{i+1} \delta_{i+1,j} \dd{t} \\
    \qty(\intc_i(t) - 1) \qty(\intc_j(t) - 1)
    & \frac{1}{2} \hata_{i+1} \delta_{i,j} \dd{t} \\
    \qty(\intc_i(t) + 1) \qty(\intc_j(t) + 1)
    & \frac{1}{2} \hata_{i} \delta_{i,j} \dd{t} \\
    \qty(\intc_i(t) + 1) \qty(\intc_j(t) - 1)
    & \frac{1}{2} \hata_{i} \delta_{i-1,j} \dd{t} \\
    \qty(\intc_i(t) - 1) \intc_j(t)
    & \frac{1}{2} \hata_{i+1} (1 - \delta_{i+1,j} - \delta_{i,j}) \dd{t}
    \\
    \qty(\intc_i(t) + 1) \intc_j(t)
    & \frac{1}{2} \hata_{i} (1 - \delta_{i-1,j} - \delta_{i,j}) \dd{t} \\
    \intc_i(t) \qty(\intc_j(t) - 1) 
    & \frac{1}{2} \hata_{j+1} (1 - \delta_{i-1,j} - \delta_{i,j}) \dd{t}
    \\
    \intc_i(t) \qty(\intc_j(t) + 1) 
    & \frac{1}{2} \hata_{j} (1 - \delta_{i+1,j} - \delta_{i,j}) \dd{t} \\
    \qty(\intc_i(t) + 2) \qty(\intc_j(t) + 2)
    & \frac{1}{4} \hata_i \delta_{i,j} \dd{t} \\
    \qty(\intc_i(t) - 2) \qty(\intc_j(t) - 2)
    & \frac{1}{4} \hata_{i+1} \delta_{i,j} \dd{t} \\
    \qty(\intc_i(t) + 2) \intc_j(t)
    & \frac{1}{4} \hata_i \qty(1 - \delta_{i,j}) \dd{t} \\
    \intc_i(t) \qty(\intc_j(t) +2)
    & \frac{1}{4} \hata_j \qty(1 - \delta_{i,j}) \dd{t} \\
    \qty(\intc_i(t) - 2) \intc_j(t)
    & \frac{1}{4} \hata_{i+1} \qty(1 - \delta_{i,j}) \dd{t} \\
    \intc_i(t) \qty(\intc_j(t) - 2)
    & \frac{1}{4} \hata_{j+1} \qty(1 - \delta_{i,j}) \dd{t} \\
    \intc_i(t) \intc_j(t)
    & \qty( 1 - \Sigma \dd{t}),
  \end{cases}
\end{equation}
where $\Sigma \dd{t}$ is the probability of happening all the events,
mentioned in the update rules in the 
time-interval $t$ and $t+\dd{t}$.
Using this update rules and eq.\eqref{eq:current approximation}, we
can write the evolution equation 
of the equal-time correlation function of current as
\begin{align}
  \label{eq:equal time intc intc evl eqn}
  \dv{t} C_r^{\intc \intc}(t,t)
  =& \Gamma_r(t) +
     \angular{\qty[\hata_0(t) - \hata_1(t)] \intc_r(t)}_c +
     \angular{\intc_0 \qty[\hata_r - \hata_{r+1}]}_c
     \nonumber \\ \simeq
   & \Gamma_r(t) +
     a^\prime\qty[c^{m \intc}_r(t,t) - c^{m \intc}_{r-1}(t,t)] +
     a^\prime\qty[c^{m \intc}_{L-r}(t,t) - c^{m \intc}_{L-r-1}(t,t)].
\end{align}
where,
\begin{align}
  \label{eq:gammar def with spatial index}
  \Gamma_r \equiv \Gamma_{i,j} = \frac{3}{2} \delta_{i,j}
  \angular{\hata_{i+1} + 
  \hata_i} - \frac{1}{2} \delta_{i+1,j} \angular{\hata_{i+1}} -
  \frac{1}{2} \delta_{i-1,j} \angular{\hata_i}.
\end{align}
The solution of eq.\eqref{eq:equal time intc intc evl eqn}
appeared in eq.\eqref{eq:same time curr corr fourier} in the
main-text. 

\subsection{Mass and integrated current correlation}
\label{sec:mass intc corr evl eqn}

The update rules of the temporal evolution equation of the product
function of mass and integrated current, are given as
\begin{align}
  \label{eq:equal time mass curr update rules}
  m_i(t+\dd{t})\intc_j(t+\dd{t}) =
  \begin{cases}
    \qty(m_i(t) + 1) \qty(\intc_j(t) -1)
    & \frac{1}{2} \hata_{i+1} \delta_{i,j} \dd{t} \\
    \qty(m_i(t) + 1) \qty(\intc_j(t) +1)
    & \frac{1}{2} \hata_{i+1} \delta_{i+1,j} \dd{t} \\
    \qty(m_i(t) - 2) \qty(\intc_j(t) -1)
    & \frac{1}{2} \hata_{i} \delta_{i-1,j} \dd{t} \\
    \qty(m_i(t) - 2) \qty(\intc_j(t) + 1)
    & \frac{1}{2} \hata_{i} \delta_{i,j} \dd{t} \\
    \qty(m_i(t) + 1) \qty(\intc_j(t) -1)
    & \frac{1}{2} \hata_{i-1} \delta_{i-2,j} \dd{t} \\
    \qty(m_i(t) + 1) \qty(\intc_j(t) +1)
    & \frac{1}{2} \hata_{i-1} \delta_{i-1,j} \dd{t} \\
    \qty(m_i(t) - 2) \qty(\intc_j(t) + 2)
    & \frac{1}{4} \hata_i \delta_{i,j} \dd{t} \\
    \qty(m_i(t) + 2) \qty(\intc_j(t) + 2)
    & \frac{1}{4} \hata_{i-1} \delta_{i-1,j} \dd{t} \\
        \qty(m_i(t) + 2) \qty(\intc_j(t) - 2)
    & \frac{1}{4} \hata_{i+1} \delta_{i,j} \dd{t} \\
    \qty(m_i(t) - 2) \qty(\intc_j(t) - 2)
    & \frac{1}{4} \hata_{i} \delta_{i-1,j} \dd{t} \\
    \qty(m_i(t) + 1) \intc_j(t)
    & \frac{1}{2} \hata_{i+1} \qty(1 - \delta_{i,j} - \delta_{i+1,j})
      \dd{t} \\
    \qty(m_i(t) + 1) \intc_j(t)
    & \frac{1}{2} \hata_{i-1} \qty(1 - \delta_{i-1,j} - \delta_{i-2,j})
      \dd{t} \\
    \qty(m_i(t) - 2) \intc_j(t)
    & \frac{1}{2} \hata_{i} \qty(1 - \delta_{i,j} - \delta_{i-1,j})
      \dd{t} \\
    \qty(m_i(t) + 2) \intc_j(t)
    & \frac{1}{4} \hata_{i+1} \qty(1 - \delta_{i,j})
      \dd{t} \\
    \qty(m_i(t) - 2) \intc_j(t)
    & \frac{1}{4} \hata_{i} \qty(1 - \delta_{i-1,j})
      \dd{t} \\
    \qty(m_i(t) - 2) \intc_j(t)
    & \frac{1}{4} \hata_{i} \qty(1 - \delta_{i,j})
      \dd{t} \\
    \qty(m_i(t) + 2) \intc_j(t)
    & \frac{1}{4} \hata_{i-1} \qty(1 - \delta_{i-1,j})
      \dd{t} \\
    m_i(t) \qty(\intc_j(t) + 1)
    & \frac{1}{2} \hata_j \qty(1 - \delta_{i,j} - \delta_{i,j-1} -
      \delta_{i,j+1}) \dd{t} \\
    m_i(t) \qty(\intc_j(t) - 1)
    & \frac{1}{2} \hata_{j+1} \qty(1 - \delta_{i,j} - \delta_{i,j+1} -
      \delta_{i,j+2}) \dd{t} \\
    m_i(t) \qty(\intc_j(t) + 2)
    & \frac{1}{4} \hata_j \qty(1 - \delta_{i,j} - \delta_{i,j+1})
      \dd{t} \\
    m_i(t) \qty(\intc_j(t) - 2)
    & \frac{1}{4} \hata_{j+1} \qty(1 - \delta_{i,j} - \delta_{i,j+1})
      \dd{t}.
  \end{cases}
\end{align}
Using these  rules, we write the evolution equation of the two
point mass and integrated current correlation as
\begin{align}
  \label{eq:mass intc corr evl eqn}
  \dv{t} C_r^{m \intc} (t,t) = f_r(t) +
  \sum\limits_k \Delta_{r,k} C_k^{\hata \intc} (t,t),
\end{align}
where $f_r(t)$, the source term of the equal time
mass-integrated current correlation, which, in the steady state can be
written as
\begin{align}
  \label{eq:mass intc source term}
  f_r(t)  = \qty[C^{m \hata}_r(t,t) - C^{m \hata}_{r+1} (t,t)] +
  \frac{7a(\rhobar)}{2} \delta_{0,r+1} -
  \frac{7a(\rhobar)}{2} \delta_{0,r} + \frac{a(\rhobar)}{2}
  \qty(\delta_{0,r-1} - \delta_{0,r+2}).
\end{align}
The Fourier transform of eq.~\eqref{eq:mass intc corr evl eqn} is used
to get eq.\eqref{eq:evl eqn same time mQ corr} in the main-text.

\subsection{Mass-mass correlation function}
\label{sec:mass mass correlation}
\label{sec:appendix mass activity correlation}

The equal time mass-mass correlation function is important to
calculate the mass-activity correlation function and the power
spectrum of subsystem mass. The update rules are the following,
\begin{align}
  \label{eq:mass mass corr update eqn}
  m_i(t+\dd{t}) m_j(t+\dd{t}) =
  \begin{cases}
    \qty(m_i(t) + 1) \qty(m_j(t) + 1)
    & \frac{1}{2} \hata_{i+1} \delta_{i,j} \dd{t} \\
    \qty(m_i(t) + 1) \qty(m_j(t) - 2)
    & \frac{1}{2} \hata_{i+1} \delta_{i+1,j} \dd{t} \\
    \qty(m_i(t) + 1) \qty(m_j(t) + 1)
    & \frac{1}{2} \hata_{i+1} \delta_{i+2,j} \dd{t} \\
    \qty(m_i(t) - 2) \qty(m_j(t) + 1)
    & \frac{1}{2} \hata_i \delta_{i-1,j} \dd{t} \\
    \qty(m_i(t) - 2) \qty(m_j(t) - 2)
    & \frac{1}{2} \hata_i \delta_{i,j} \dd{t} \\
    \qty(m_i(t) - 2) \qty(m_j(t) + 1)
    & \frac{1}{2} \hata_i \delta_{i+1,j} \dd{t} \\
    \qty(m_i(t) + 1) \qty(m_j(t) + 1)
    & \frac{1}{2} \hata_{i-1} \delta_{i-2,j} \dd{t} \\
    \qty(m_i(t) + 1) \qty(m_j(t) - 2)
    & \frac{1}{2} \hata_{i-1} \delta_{i-1,j} \dd{t} \\
    \qty(m_i(t) + 1) \qty(m_j(t) + 1)
    & \frac{1}{2} \hata_{i-1} \delta_{i,j} \dd{t} \\
    \qty(m_i(t) + 1) m_j(t)
    & \frac{1}{2} \hata_{i+1} \qty(1 - \delta_{i,j} - \delta_{i+1,j} -
      \delta_{i+2,j}) \dd{t} \\
    \qty(m_i(t) - 2) m_j(t)
    & \frac{1}{2} \hata_{i} \qty(1 - \delta_{i-1,j} - \delta_{i,j} -
      \delta_{i+1,j}) \dd{t} \\
    \qty(m_i(t) + 1) m_j(t)
    & \frac{1}{2} \hata_{i-1} \qty(1 - \delta_{i-2,j} - \delta_{i-1,j} -
      \delta_{i,j}) \dd{t} \\
    m_i(t) \qty(m_j(t) + 1)
    & \frac{1}{2} \hata_{j+1} \qty(1 - \delta_{i,j} - \delta_{i,j+1}
      -\delta_{i,j+2}) \dd{t} \\
    m_i(t) \qty(m_j(t) - 2)
    & \frac{1}{2} \hata_{j} \qty(1 - \delta_{i,j-1} - \delta_{i,j}
      -\delta_{i,j+1}) \dd{t} \\
    m_i(t) \qty(m_j(t) + 1)
    & \frac{1}{2} \hata_{j-1} \qty(1 - \delta_{i,j-2} - \delta_{i,j-1}
      -\delta_{i,j}) \dd{t} \\
    \qty(m_i(t) + 2)\qty(m_j(t) + 2)
    & \frac{1}{4} \hata_{i+1} \delta_{i,j} \dd{t} \\
    \qty(m_i(t) + 2)\qty(m_j(t) - 2)
    & \frac{1}{4} \hata_{i+1} \delta_{i+1,j} \dd{t} \\
    \qty(m_i(t) - 2)\qty(m_j(t) + 2)
    & \frac{1}{4} \hata_{i} \delta_{i-1,j} \dd{t} \\
    \qty(m_i(t) - 2)\qty(m_j(t) - 2)
    & \frac{1}{4} \hata_{i} \delta_{i,j} \dd{t} \\
    \qty(m_i(t) + 2)m_j(t)
    & \frac{1}{4} \hata_{i+1} \qty(1 - \delta_{i,j} - \delta_{i+1,j})
      \dd{t} \\
    \qty(m_i(t) - 2)m_j(t)
    & \frac{1}{4} \hata_{i} \qty(1 - \delta_{i-1,j} - \delta_{i,j})
      \dd{t} \\
      m_i(t) \qty(m_j(t) + 2)
    & \frac{1}{4} \hata_{j+1} \qty(1 - \delta_{i,j} - \delta_{i,j+1})
      \dd{t} \\
    m_i(t) \qty(m_j(t) - 2)
    & \frac{1}{4} \hata_{j} \qty(1 - \delta_{i,j-1} - \delta_{i,j})
      \dd{t} \\
    \qty(m_i(t) - 2)\qty(m_j(t) - 2)
    & \frac{1}{4} \hata_{i} \delta_{i,j} \dd{t} \\
    \qty(m_i(t) - 2)\qty(m_j(t) + 2)
    & \frac{1}{4} \hata_{i} \delta_{i+1,j} \dd{t} \\
    \qty(m_i(t) + 2)\qty(m_j(t) - 2)
    & \frac{1}{4} \hata_{i-1} \delta_{i-1,j} \dd{t} \\
    \qty(m_i(t) + 2)\qty(m_j(t) + 2)
    & \frac{1}{4} \hata_{i-1} \delta_{i,j} \dd{t} \\
    \qty(m_i(t) - 2)m_j(t)
    & \frac{1}{4} \hata_{i} \qty(1 - \delta_{i,j} - \delta_{i+1,j})
      \dd{t} \\
    \qty(m_i(t) + 2)m_j(t)
    & \frac{1}{4} \hata_{i-1} \qty(1 - \delta_{i-1,j} - \delta_{i,j})
      \dd{t} \\
      m_i(t) \qty(m_j(t) + 2)
    & \frac{1}{4} \hata_{j-1} \qty(1 - \delta_{i,j} - \delta_{i,j-1})
      \dd{t} \\
    m_i(t) \qty(m_j(t) - 2)
    & \frac{1}{4} \hata_{j} \qty(1 - \delta_{i,j+1} - \delta_{i,j})
      \dd{t}
  \end{cases}
\end{align}
From these update rules, we  can write the evolution equation of
$C^{mm}_r(t,t)$ as
\begin{align}
  \label{eq:mass mass evl eqn equal time}
  \dv{t} \angular{m_i(t) m_j(t)} = \sum\limits_k
  \angular{m_{i} \Delta_{j,k} \hata_k + \Delta_{i,k} \hata_k m_j}_c + B_{i,j}.
\end{align}
where
\begin{align}
  \label{eq:Bij general form}
  B_{i,j}
  =& \frac{1}{2}\angular{3\hata_{i-1} + 8\hata_{i} +
     3\hata_{i+1}} \delta_{i,j} \nonumber \\
   & -\frac{1}{2}\angular{4\hata_{i+1} + 4\hata_i}\delta_{i+1,j}
     -\frac{1}{2}\angular{4\hata_{i-1} + 4\hata_i}\delta_{i+1,j}
     \nonumber \\
   &     + \frac{1}{2}\angular{\hata_{i+1}}\delta_{i+2,j}
     + \frac{1}{2}\angular{\hata_{i-1}}\delta_{i-2,j},
\end{align}
is the source of the equal-time mass correlation. In the steady state,
where $\angular{\hata_i(t)} = a(\rhobar)$, we can write $B_{i,j}$
as a translationally invariant form,
$B_{i,j} \equiv B_r$ as
\begin{align}
  B_{r} =  7a(\rhobar)\delta_{0,r} -4a(\rhobar)
  \qty(\delta_{0,r+1} + \delta_{0,r-1})+ 
  \frac{a(\rhobar)}{2}\qty(\delta_{0,r+2} + \delta_{0,r-2}).
\end{align}
Equation~\eqref{eq:mass mass evl eqn equal time} has appeared in 
eq.\eqref{eq:mass activity evl eqn same time} in the main-text.
Using the steady state condition, we must have
$d \angular{m_i m_j}_c/dt = 0$, which implies,
\begin{align}
  \label{eq:steady state mm corr}
  2\qty[C^{m\hata}_{r-1} - 2C^{m\hata}_{r} + C^{m\hata}_{r+1}] + B_r = 0.
\end{align}
Eq.\eqref{eq:steady state mm corr} can be solved by considering the
following generating function,
\begin{align}
  \label{eq:generating function def}
  G\qty(z) = \sum\limits_{r = 0}^{\infty} C^{m\hata}_{r} z^r.
\end{align}
We multiply both side of eq.\eqref{eq:steady state mm corr} with $z^r$
and sum over $r$ to get,
\begin{align}
  \label{eq:generating function manna}
  G(z) = \frac{4 C^{m\hata}_{0} - 4 z C^{m\hata}_{1}-z a \qty[(z-8) z+14]}
  {4 (1-z)^2},
\end{align}
where we use the identities,
\begin{align}
 \sum\limits_{r=0}^{\infty} C^{m\hata}_{r-1} z^r &= C^{m\hata}_{1} + z
  G\qty(z), \nonumber \\
  \sum\limits_{r=0}^{\infty} C^{m\hata}_{r+1} z^r &=
  \frac{G(z) - C^{m\hata}_{0} }{z}.
\end{align}
As we are dealing with truncated correlation functions, in the limit
$z \rightarrow 1$, we must have $\lim_{z \rightarrow 1} G(z) <
\infty$. Using a new
variable  $w \rightarrow 1-z$, we write
eq.\eqref{eq:generating function manna} as
\begin{align}
  \label{eq:transformed generating function}
  G(w) = \frac{1}{4w^2}
  \qty[a(\rhobar) w^3 + 5 a(\rhobar) w^2 +(a(\rhobar) + 4 C^{m\hata}_{1})
  w + -7 a(\rhobar) + 4 C^{m\hata}_{0} - 4 C^{m\hata}_{1}]
\end{align}
For the convergence of $G(w)$ in the limit $w \rightarrow 0$, we set
\begin{align}
  \label{eq:conditions}
  a(\rhobar) + 4 C^{m\hata}_{1} = 0;
  -7 a(\rhobar) + 4 C^{m\hata}_{0} - 4 C^{m\hata}_{1} = 0,
\end{align}
leading to the following exact relations,
\begin{align}
  \label{eq:final mm corr sol}
  C^{m\hata}_{0} = \frac{3 a(\rhobar)}{2}, \\
  C^{m\hata}_{1} = -\frac{a(\rhobar)}{4}.
\end{align}
Finally, putting eq.\eqref{eq:final mm corr sol} in
eq.\eqref{eq:generating function manna} we get the generating function
\begin{align}
  \label{eq:final manna generating function}
  G\qty(z) = \frac{3 a(\rhobar)}{2} - \frac{a(\rhobar)}{4} z.
\end{align}

\end{widetext} 
\end{document}